\documentclass[twocolumn]{aastex701}
\usepackage{amsmath}
\usepackage{hyperref}
\usepackage{cleveref}
\usepackage{placeins}
\usepackage{float}
\newcommand{\chng}[1]{#1}
\shorttitle{GeV Emission Search in Magnetar Giant Flares}
\shortauthors{Vikas Chand et al.}

\newcommand{\fermi}{\textit{Fermi}}
\newcommand{\lat}{{LAT}}

\begin{document}

\title{A Search for GeV Emission from Magnetar Giant Flare Candidates with \fermi-LAT}

\author[0000-0002-7876-7362]{Vikas Chand}
\affiliation{Department of Physics \& Astronomy, Louisiana State University, Baton Rouge, LA 70803, USA}
\email{Vikas.Chand@lsu.edu}

\author[0000-0002-6548-5622]{Michela Negro}
\affiliation{Department of Physics \& Astronomy, Louisiana State University, Baton Rouge, LA 70803, USA}
\email{m@lsu.edu}

\author[0000-0002-5448-7577]{Nicola Omodei}
\affiliation{W. W. Hansen Experimental Physics Laboratory, Kavli Institute for Particle Astrophysics and Cosmology, Department of Physics and SLAC National Accelerator Laboratory, Stanford University, Stanford, CA 94305, USA}
\email{n@lsu.edu}

\author[0000-0002-7574-1298]{Niccol\`o Di Lalla}
\affiliation{W. W. Hansen Experimental Physics Laboratory, Kavli Institute for Particle Astrophysics and Cosmology, Department of Physics and SLAC National Accelerator Laboratory, Stanford University, Stanford, CA 94305, USA}
\email{n@lsu.edu}

\author[0000-0002-2942-3379]{Eric Burns}
\affiliation{Department of Physics \& Astronomy, Louisiana State University, Baton Rouge, LA 70803, USA}
\email{n@lsu.edu}

\author[0000-0002-0130-2460]{Soebur Razzaque}
\affiliation{University of Johannesburg: Auckland Park, Gauteng, ZA}
\email{n@lsu.edu}

\author[0009-0006-8598-728X]{Aaron C. Trigg}
\affiliation{Department of Physics \& Astronomy, Louisiana State University, Baton Rouge, LA 70803, USA}
\email{n@lsu.edu}

\author[0000-0003-0406-7387]{Giacomo Principe}
\affiliation{Dipartimento di Fisica, Universitá di Trieste, I-34127 Trieste, Italy}
\email{n@lsu.edu}

\begin{abstract}
The discovery of delayed GeV emission from the extragalactic magnetar giant flare (MGF) GRB\,200415A, located in the nearby Sculptor galaxy, revealed for the first time that these rare transients can launch relativistic outflows that power high-energy afterglows.
Motivated by the recent identification of additional nearby MGF candidates in the archival data of the \textit{Fermi} Gamma-ray Burst Monitor, we conduct a search for GeV counterparts with the \textit{Fermi} Large Area Telescope (LAT). We analysed post-trigger time intervals taken in the range $10^{2}$--$10^{4}$\,s using a maximum-likelihood approach and performed a stacking analysis of all candidates with LAT coverage. In addition, we searched for photon triplets through \chng{a waiting time} analysis to identify events potentially associated with MGFs.
We recover the known delayed signal from GRB\,200415A but find no GeV emission from the remaining six candidates. For three events, the earliest emission is unconstrained because the $10^{2}$\,s interval contains zero exposure after standard selections. For the events with LAT coverage, we obtain upper limits at 95\% confidence level 
of order $F_E \sim 10^{-9}\,\mathrm{erg\,cm^{-2}\,s^{-1}}$ 
for the individual events, and a stacked population-averaged limit of $\approx2\times10^{-10}\,\mathrm{erg\,cm^{-2}\,s^{-1}}$. 
We interpret these upper limits within the 
relativistic fireball framework, where the prompt 
spectral peaks favor a baryonic-poor regime 
($\eta > \eta_*$). For the candidates 
with hard prompt spectrum and early LAT coverage, 
the baryonic-poor condition restricts the mass of relativistic ejecta 
to $M_b \lesssim 1\text{--}4 \times 10^{22}$\,g; 
the LAT upper limits confirm that the predicted GeV 
afterglow from such clean outflows falls below 
current instrumental sensitivity.
\end{abstract}


\section{Introduction}\label{sec:intro}
Magnetars are neutron stars that exhibit both transient  soft gamma-ray activity and persistent X-ray emission. Their emission is driven by the evolution and decay of ultrastrong magnetic fields ($B \gtrsim 5\times10^{13}$--$10^{15}$G) \citep[e.g.,][for a review]{Kaspi:2017, Negro:2024FrASS1188953N}. While they typically exhibit persistent X-ray emission ($L_{\rm X} \sim 10^{33}$--$10^{36}$~erg~s$^{-1}$), long-period brightening (outbursts) and short bursts, their most extreme manifestations are magnetar giant flares (MGFs). These events release energies of the order of $10^{44}$–$10^{47}$~erg within a fraction of a second, characterized by a hard, non-thermal initial spike followed by a softer, pulsating tail modulated at the stellar spin period \citep{Mazets:1979Natur, Hurley:1999Natur, Palmer:2005Natur}.

The interaction of MGF ejecta with the surrounding medium is expected to drive shocks capable of particle acceleration. Radio observations of the 2004 giant flare from SGR~1806$-$20 revealed an expanding nebula consistent with a mildly relativistic outflow ($\Gamma \sim 1.5$) and significant baryon loading \citep{Gaensler:2005Natur, Gelfand:2005ApJ}. However, theoretical models suggest that MGFs may also launch ultra-relativistic outflows ($\Gamma \gg 10$) analogous to those in short gamma-ray bursts (GRBs), which would generate high-energy ($>100$\,MeV) emission via synchrotron or inverse-Compton processes at the external shock \citep{Granot:2006ApJ, Lyutikov:2006MNRAS}. 

The identification of GRB~200415A, spatially coincident with the galaxy NGC~253 ($D \approx 3.7$\,Mpc), as an extragalactic MGF established a new observational class \citep{Yang:2020ApJ, Svinkin:2021Natur}. Crucially, subsequent LAT analysis revealed delayed ($\sim$19–380~s) GeV emission with a hard spectrum ($\Gamma_{\rm LAT}\simeq-1.7$), consistent with synchrotron radiation from an external shock or a bow shock driven by relativistic ejecta \citep{Zhang:2020ApJ, Ajello:2021, Chand:2021RAA}. This discovery provided the first confirmed GeV counterpart to a magnetar giant flare and established a framework for similar searches.

This GeV emission can offer a direct probe to the baryonic content in the outflow. In the ``fireball'' model, the presence of a prompt spike with a quasi-thermal spectrum requires a ``baryonic-poor" outflow (high dimensionless entropy $\eta$) that follows the $E_{\rm p} \propto E_{\rm iso}^{1/4}$ relation. On the other hand, the generation of a bright external shock afterglow requires sufficient kinetic energy, which scales with the baryonic mass \citep{Ioka:2005ApJ, Nakar:2005ApJ}. Observational constraints on the GeV flux thus define the allowable phase space for the ejecta mass and Lorentz factor, distinguishing between baryonic-rich and baryonic-poor regimes \citep{Zhang:2020ApJ}.

Recently, \citet{Trigg:2025} systematically re-analyzed the entire \textit{Fermi} Gamma-ray Burst Monitor (GBM) \citep{Meegan:2009} short-GRB catalog to identify previously overlooked extragalactic MGF candidates. They uncovered four new events, GRBs~081213A, 120616A, 200423A, and 231024A, consistent with the known MGF population in temporal and spectral properties. Together with the three previously identified events (GRB~200415A, GRB~180128A, 
and GRB~231115A), these findings expand the total extragalactic MGF sample to ten events, which includes three identifications from before the launch of \textit{Fermi}. This newly identified sample of nearby extragalactic MGFs motivates dedicated high-energy searches in the \textit{Fermi} Large Area Telescope (LAT) \citep{Atwood:2009ApJ} data to probe whether delayed GeV emission, like that observed from GRB~200415A, is a common feature of these flares.

In this paper, we present a systematic search for GeV emission from this expanded sample using \textit{Fermi}-LAT data. We perform unbinned likelihood analyses and stacked-likelihood, photon triplet analyses to probe both individual and cumulative signals to constrain the high-energy properties of these flares. We outline our data analysis in Section \ref{sec:methods}, present the search results in Section~\ref{sec:results}, and discuss the implications for the baryon loading and energetics of MGF outflows in Section~\ref{sec:discussion} and draw our conclusions in Section \ref{sec:conclusions}.

\begin{deluxetable*}{lllcccccccc}
\tabletypesize{\footnotesize}
\tablecaption{Nearby MGF candidates: GBM triggers, host galaxies, and prompt emission properties.\label{tab:mgf_sample_meta}}
\tablehead{
  \colhead{GRB} & \colhead{GBM trigger} & \colhead{Host galaxy} &
  \colhead{RA$_{\rm gal}$} & \colhead{Dec$_{\rm gal}$} &
  \colhead{$d$} &
  \colhead{$T_{90}$} & \colhead{$E_{\rm p}$} &
  \colhead{Index$^{\alpha}$} &
  \colhead{$E_{\gamma,45}$} &
  \colhead{Ref.} \\
  \colhead{} & \colhead{ID} & \colhead{} &
  \colhead{(deg)} & \colhead{(deg)} &
  \colhead{(Mpc)} &
  \colhead{(s)} & \colhead{(keV)} &
  \colhead{} &
  \colhead{($10^{45}$ erg)} &
  \colhead{}
}
\startdata
GRB 081213A  & bn081213173 & NGC 253  &  11.8881 & $-25.2888$ & 3.7 & 0.050 &  400 & $-0.6$ & 0.42 & 1 \\
GRB 180128A  & bn180128215 & NGC 253  &  11.8881 & $-25.2888$ & 3.7 & 0.155 &  290 & $-0.6$ & 0.60 & 2 \\
GRB 200415A  & bn200415367 & NGC 253  &  11.8881 & $-25.2888$ & 3.7 & 0.200 & 1080 & $-0.0$ & 14.2 & 3 \\
GRB 231024A  & bn231024556 & NGC 253  &  11.8881 & $-25.2888$ & 3.7 & 0.094 &  500 & $-0.8$ & 0.55 & 4 \\
\hline
GRB 120616A  & bn120616630 & IC 342   &  56.7021 & $+68.0961$ & 2.3 & 0.050 &  500 & $-0.4$ & 0.23 & 4 \\
GRB 200423A  & bn200423579 & NGC 6946 & 308.7181 & $+60.1537$ & 7.7 & 0.032 &  600 & $-0.7$ & 8.50 & 4 \\
GRB 231115A  & bn231115650 & M82      & 148.9685 & $+69.6797$ & 3.5 & 0.097 &  600 & $-0.1$ & 1.15 & 5 \\
\enddata
\tablecomments{
$E_{\gamma,45} \equiv E_{\gamma,\mathrm{iso}}/10^{45}\,\mathrm{erg}$.
Distances are approximate values for the host galaxies.
}
\tablerefs{
(1)~\citet{Bissaldi:2008GCN};
(2)~\citet{Trigg:2024};
(3)~\citet{Ajello:2021};
(4)~\citet{Trigg:2025};
(5)~\citet{Trigg:2025M82}
}
\end{deluxetable*}

\begin{deluxetable*}{l c c c c c l}
\tablecaption{Fermi--LAT Analysis Results \label{tab:lat_results}}
\tablewidth{0pt}
\tablehead{
\colhead{GRB} &
\colhead{Time window (s)} &
\colhead{TS} &
\colhead{Photon flux} &
\colhead{Energy flux} &
\colhead{Index} &
\colhead{Notes} \\
\colhead{} & \colhead{} & \colhead{} &
\colhead{[$10^{-6}$\,ph\,cm$^{-2}$\,s$^{-1}$]} &
\colhead{[$10^{-9}$\,erg\,cm$^{-2}$\,s$^{-1}$]} &
\colhead{} & \colhead{}
}
\startdata
\multicolumn{7}{l}{GRB\,081213A} \\
 & $10^{2}$  & 1   & $<11.1$  & $<8.27$  & $-2^{\dagger}$ & No detection (GTI $\sim10^{2}$\,s) \\
 & $10^{3}$  & 0   & $<3.69$  & $<2.75$  & $-2^{\dagger}$ & No detection (GTI $\sim$281\,s) \\
 & $10^{4}$  & 0   & $<0.548$ & $<0.408$ & $-2^{\dagger}$ & No detection (GTI $\sim$2.84\,ks) \\ \hline
 \multicolumn{7}{l}{GRB\,180128A} \\
 & $10^{2}$  & --- & ---     & ---      & ---             & \chng{Exposure $=0$} \\ \\
  & $10^{3}$  & 0   & $<1.13$  & $<0.85$  & $-2^{\dagger}$  & No detection; first exposure at $\sim$148\,s \\
 & $10^{4}$  & 0   & $<0.315$ & $<0.234$ & $-2^{\dagger}$  & No detection (GTI $\sim$5.52\,ks; first exposure at $\sim$148\,s) \\ \hline
 \multicolumn{7}{l}{GRB\,200415A} \\
 & $10^{2}$  & 10 & \chng{$<12.6$} & $<9.42$ & $-2^{\dagger}$ &  Mild excess\\
 & $10^{3}$  & \textbf{25} & $3.60\pm2.00$ & $4.14\pm2.33$ & $-1.71\pm0.34$ & LAT detection ($\sim$5$\sigma$); hard spectrum \\
 & $10^{4}$  & 22 & $1.62\pm0.96$ & $1.97\pm1.16$ & $-1.67\pm0.36$ & Detection (GTI $\sim$4.17\,ks) \\
\hline
\multicolumn{7}{l}{GRB\,231115A} \\
 & $10^{2}$  & 3   & $<9.28$  & $<6.92$  & $-2^{\dagger}$ & No detection (low TS)\\
 & $10^{3}$  & 1   & $<1.40$  & $<1.05$  & $-2^{\dagger}$ & No detection \\
 & $10^{4}$  & 1   & $<0.95$  & $<0.707$  & $-2^{\dagger}$ & No detection (exposure $\sim$6.1\,ks) \\
\hline\hline
\multicolumn{7}{l}{GRB\,120616A} \\
 & $10^{2}$  & --- & --- & --- & --- & ROI outside LAT FoV; exposure $=0$ \\
 & $10^{3}$  & --- & --- & --- & --- & ROI outside LAT FoV; exposure $=0$ \\
 & $10^{4}$  & 1   & $<0.67$ & $<0.50$ & $-2^{\dagger}$ & No detection (exposure $\sim$4.2\,ks; first exposure at $\sim$1447\,s) \\
\hline
\multicolumn{7}{l}{GRB\,200423A} \\
 & $10^{2}$  & --- & --- & --- & --- & ROI outside LAT FoV; exposure $=0$ \\
 & $10^{3}$  & 1   & $<2.67$ & $<1.99$ & $-2^{\dagger}$ & No detection; first exposure at $\sim$619\,s \\
 & $10^{4}$  & 0   & $<0.34$ & $<0.25$ & $-2^{\dagger}$ & No detection; first exposure at $\sim$619\,s \\
\hline
\multicolumn{7}{l}{GRB\,231024A} \\
 & $10^{2}$  & 0   & $<73.1$ & $<54.5$ & $-2^{\dagger}$ & No detection (no events in $10^{2}$\,s ROI) \\
 & $10^{3}$  & 0   & $<1.74$ & $<1.29$ & $-2^{\dagger}$ & No detection (GTI $\sim10^{3}$\,s) \\
 & $10^{4}$  & 0   & $<0.37$ & $<0.28$ & $-2^{\dagger}$ & No detection \\
\enddata
\tablecomments{Fluxes refer to $0.1$--$10\,\mathrm{GeV}$. ULs are 95\% C.L. profile-likelihood values with the photon index fixed to $-2$ (denoted $^{\dagger}$)
}
\end{deluxetable*}

\begin{deluxetable*}{lc|cccc|ccc}
\tablecaption{Photon Triplet Analysis Results\label{tab:triplet_results}}
\tablewidth{0pt}
\tablehead{
\colhead{Source} & \colhead{Host} &
\multicolumn{4}{c}{Next Triplet Analysis} &
\multicolumn{3}{c}{Next-Best Triplet Analysis} \\
\cline{3-6} \cline{7-9}
\colhead{} & \colhead{} &
\colhead{$\sigma$} & \colhead{$p$} & \colhead{$\Delta t$} & \colhead{Delay} &
\colhead{$\Delta t$} & \colhead{Delay} & \colhead{\chng{$p$ ($\sigma$)}} \\
\colhead{} & \colhead{} &
\colhead{} & \colhead{} & \colhead{(s)} & \colhead{(s)} &
\colhead{(s)} & \colhead{(days)} & \colhead{}
}
\startdata
GRB\,200415A & NGC\,253 & 5.39 & $3.8\times10^{-5}$ & 265 & 20 & \chng{265} & \chng{$\sim$0.0} & \chng{$3.8\times10^{-5}$ (3.96)\tablenotemark{a}} \\
GRB\,231115A & M82 & 0.81 & 0.85 & 81,340 & 1,406 & \chng{458} & \chng{208.8} & \chng{0.34 (0.4)} \\
GRB\,200423A & NGC\,6946 & 0.58 & 0.34 & 16,153 & 7,554 & \chng{154} & \chng{209.7} & \chng{0.46 (0.1)} \\
GRB\,180128A & NGC\,253 & 0.41 & 0.63 & 114,160 & 35,479 & \chng{482} & \chng{332.2} & \chng{0.086 (1.4)} \\
GRB\,231024A & NGC\,253 & 0.22 & 0.39 & 103,068 & 115,840 & \chng{119} & \chng{40.5} & \chng{0.005 (2.6)\tablenotemark{b}} \\
GRB\,081213A & NGC\,253 & 0.14 & 0.33 & 84,045 & 63,965 & \chng{595} & \chng{211.8} & \chng{0.069 (1.5)} \\
GRB\,120616A & IC\,342 & 0.11 & 0.80 & 20,557 & 13,262 & \chng{85} & \chng{355.8} & \chng{0.27 (0.6)} \\
\enddata
\tablecomments{
Photon triplet analysis using \textit{Fermi}-LAT data (0.1--100~GeV) within a $1^{\circ}$ ROI.
\textit{The next triplet Analysis} (columns 3--6): Li\&Ma significance ($\sigma$) \chng{and the analytic waiting-time false-alarm probability ($p = F(\Delta t)$), which are distinct statistics of the same triplet},
triplet time interval ($\Delta t = t_3 - t_1$), and delay from trigger to first photon for the first triplet
following each trigger.
\chng{\textit{The next-best triplet analysis} (columns 7--9): the most compact (minimum-$\Delta t$) triplet
within each one-year post-trigger window, its delay from trigger, and its
per-window trials-corrected false-alarm probability $p$ (with the equivalent Gaussian significance
$\sigma$ in parentheses) from a $10^{5}$ realization of  photon-time scramble
(Figure~\ref{fig:mgf_triplet_demographics}).} Only GRB\,200415A shows significant prompt GeV emission ($\sigma = 5.39$, $p = 3.8\times10^{-5}$,
delay $\sim$20~s).}
\tablenotetext{a}{\chng{The next-best triplet of GRB\,200415A is its pre-specified prompt triplet; the
quoted $p = 3.8\times10^{-5}$ ($3.96\sigma$) is the analytic waiting-time value of corresponding to that  triplet.}}
\tablenotetext{b}{\chng{GRB\,231024A is the most significant next-best triplet ($p = 0.005$, $2.6\sigma$);
after a Bonferroni correction over the six non-detected events its significance decreases to $1.9\sigma$.}}
\end{deluxetable*}

\begin{deluxetable*}{lcccccccc}
\tabletypesize{\footnotesize}
\tablecaption{Derived fireball parameters and constraints on the baryonic outflow mass.\label{tab:fireball_baryon}}
\tablehead{
  \colhead{GRB} &
  \colhead{$L_{\gamma,47}$} &
  \colhead{$E_{\mathrm{LAT,iso}}$} &
  \colhead{$T_0$} &
  \colhead{$\eta_*$} &
  \colhead{$M_{b,22}^{\rm (BP)}$} &
  \colhead{$M_{b,22}^{\rm (LAT)}$} &
  \colhead{\chng{$E_{\mathrm{k,iso}}/E_{\mathrm{k,iso}}(\mathrm{200415A})$}} \\
  \colhead{} &
  \colhead{($10^{47}\,\mathrm{erg\,s^{-1}}$)} &
  \colhead{($10^{46}$ erg)} &
  \colhead{(keV)} &
  \colhead{} &
  \colhead{($10^{22}$ g)} &
  \colhead{($10^{22}$ g)} &
  \colhead{}
}
\startdata
GRB 081213A  & 0.08 & $<0.45$ & 148 &  75 & $<2.1$  & $<65.6$  & 0.03 \\
GRB 180128A  & 0.04 & $\cdots$ & 122 &  62 & $<3.6$  & $\cdots$  & 0.04 \\
GRB 200415A  & 0.71 & 0.68\tablenotemark{a} & 252 & 129 & $<40.8$ & $>58.3$\tablenotemark{a} & 1.00 \\
GRB 231024A  & 0.06 & $<0.21$ & 135 &  69 & $<3.0$  & $<33.9$  & 0.04 \\
GRB 120616A  & 0.05 & $\cdots$ & 127 &  65 & $<1.3$  & $\cdots$ & 0.02 \\
GRB 200423A  & 2.66 & $\cdots$ & 351 & 179 & $<17.6$ & $\cdots$  & 0.60 \\
GRB 231115A  & 0.12 & $<0.16$ & 161 &  82 & $<5.2$  & $<21.6$  & 0.08 \\
\enddata
\tablecomments{
$L_{\gamma,47}$ is the isotropic $\gamma$-ray luminosity in the GBM band, in units of $10^{47}\,\mathrm{erg\,s^{-1}}$, obtained using
$E_{\gamma,\mathrm{iso}} = L_{\gamma,\mathrm{iso}} T_{90}$.
The LAT constraints ($E_{\mathrm{LAT,iso}}$, $M_{b,22}^{\rm (LAT)}$) use the $10^{3}$\,s integration window (see Section~\ref{sec:discussion}) and are quoted only for events with LAT exposure from the trigger time. No LAT constraint is listed for GRBs\,180128A, 200423A, and 120616A, whose first LAT exposure begins only at $\sim$148, $\sim$619, and $\sim$1447\,s, respectively; for these events a $10^{3}$\,s isotropic-energy budget would require assuming emission during the unobserved early interval, which is precisely the epoch left unconstrained.
$E_{\mathrm{LAT,iso}}$ is the 0.1\,--10\,GeV isotropic energy, in units of $10^{46}$ erg.
$M_{b,22}^{\rm (BP)}$ is the UL on the baryonic mass from the baryonic-poor requirement $\eta > \eta_*$,
and $M_{b,22}^{\rm (LAT)}$ is the mass constraint from the LAT band, both in units of $10^{22}$ g, assuming $\xi_\gamma=0.3$ and $\xi_L=0.1$. 
$E_{\mathrm{k,iso}}/E_{\mathrm{k,iso}}(\mathrm{200415A})$ is the fiducial baryonic-rich kinetic energy normalized to GRB\,200415A; for $\xi_\gamma=0.3$ the reference value is $E_{\mathrm{k,iso}}(\mathrm{200415A}) = 3.31\times10^{46}$\,erg, and absolute values are obtained by multiplying the tabulated entries by this reference value.
\tablenotemark{a}For GRB~200415A, $E_{\mathrm{LAT,iso}}$ and $M_{b,22}^{\rm (LAT)}$ refer to the measured LAT emission and the corresponding \emph{lower} bound on the baryonic mass. Consistency with the baryonic-poor limit implies a larger effective GeV efficiency for this event, as discussed in Section~\ref{sec:discussion}.
}
\end{deluxetable*}

\section{Data and Methods}\label{sec:methods}
We analyze \fermi--LAT data for the seven GBM triggers listed in Table \ref{tab:mgf_sample_meta}, focusing on three post-trigger integration windows, $\Delta T=\{10^{2},10^{3},10^{4}\}\,$s relative to the GBM trigger time. Data were processed with the \texttt{gtburst} GUI \citep{Vianello2016gtburst}
distributed with the \texttt{Fermitools} software package 
(version 2.2.0; \citealt{fermitools}), following standard
LAT GRB methodology \citep{Ajello2019LATGRBCat2}.
We use Pass~8 (P8R3) processed data selecting transient event classes and the corresponding IRFs \citep{Atwood2013P8, Bruel2018P8}. In particular we use \texttt{P8R3\_TRANSIENT020E\_V3} event class for the shortest window and the more stringent \texttt{P8R3\_TRANSIENT010E\_V3} event class for the longer windows. In all selections we utilize FRONT+BACK event type (\texttt{evtype}=3). The maximum likelihood analysis is performed over 0.1--100~GeV, while we report photon/energy fluxes integrated over 0.1--10~GeV as described in Section \ref{subsec:upperlimits}; the highest-energy photon associated with GRB\,200415A was $\sim$1.7\,GeV \citep{Ajello:2021}, and no events above 10\,GeV are detected for any candidate, so the 0.1--10\,GeV sub-band is conservatively used to report fluxes. 
We adopt a region of interest (ROI) of radius $12^{\circ}$ centered on the MGF localization (host-galaxy coordinates). To suppress contamination from the Earth limb, a maximum zenith angle of $z_{\max}=100^{\circ}$ is applied. The sky model includes the Galactic diffuse emission template \texttt{gll\_iem\_v07.fits} and the isotropic component matching the event selection, together with all 4FGL-DR4 catalog sources within the ROI \citep{Abdollahi2022_4FGLDR3, Ballet2023_4FGLDR4}. Normalizations of sources within $5^{\circ}$ of the ROI center are left free in the fit, while parameters of more distant sources are frozen to their catalog values. Normalizations of the diffuse components are left free unless otherwise noted. Throughout this work, upper limits are computed with the photon index fixed to $\Gamma=-2$, broadly consistent with the measured power-law index of GRB\,200415A ($\Gamma\simeq-1.7\pm0.3$; \citealt{Ajello:2021}). This convention applies to individual limits (Section~\ref{subsec:upperlimits}), stacked limits, and also to the bin-by-bin spectral energy distribution (SED) scans (Section \ref{subsec:stacking}). Because the LAT field of view does not continuously 
cover a given sky position, some events have delayed 
or partial exposure within the nominal integration 
windows. In such cases, the likelihood analysis uses 
only the available good time intervals (GTIs), and 
the reported upper limits (ULs) apply to those observed
epochs; emission prior to the first exposure remains 
unconstrained.

\subsection{Detection and Upper Limits}
\label{subsec:upperlimits}
For each MGF candidate and integration window, we perform unbinned likelihood fits using \texttt{gtlike}, adopting a \texttt{PowerLaw2} spectral model that parameterizes the integrated photon flux between 0.1~GeV and 100~GeV.
The detection significance is quantified with the Test Statistic (TS), defined as ${\rm TS}=2\Delta\ln\mathcal{L}$ between models with and without the source \citep[e.g.,][]{Mattox:1996}. For each ROI, we generate TS maps both with and without the inclusion of the candidate source in the model; note that some host-galaxy positions coincide with known 4FGL sources. For these TS maps, the normalization and spectral parameters of all 4FGL sources are frozen to their catalog values, and the diffuse components are fixed to the best-fit values from the corresponding likelihood fit.

For non-detections, we derive 95\% confidence level (CL) ULs on the 0.1--10\,GeV integrated flux via profile likelihood, scanning the source flux while refitting the diffuse components and taking the flux value where $\Delta{\rm TS}=2.71$ (corresponding to the 95\% C.L. for one parameter of interest). 
In addition, for each burst we construct an SED for the first 500~s after trigger by dividing the LAT energy range into 2 logarithmic bins per decade. In each bin we refit the normalization with the index fixed to the global best-fit value (for GRB 200415A) or to the fiducial value (for non-detections), and compute ULs via the same profile-likelihood method.

\subsection{Stacking Analysis}
\label{subsec:stacking}
To search for a cumulative GeV signal from the MGF candidate population, we perform a stacked-likelihood analysis following the approach commonly used in LAT population studies \citep[e.g.,][]{Paliya:2019ApJ, Principe:2023}. For each integration window, the combined likelihood is taken as the product of the individual likelihoods for the subset of bursts with LAT exposure,
\begin{equation}
\mathcal{L}_{\rm stack}(F) = \prod_{k}\mathcal{L}_{k}(F),
\end{equation}
and the stacked test statistic is defined as
\begin{equation}
{\rm TS}_{\rm stack} = 2 \left[\ln\mathcal{L}_{\rm stack}(F)-\ln\mathcal{L}_{\rm stack}(0)\right].
\end{equation}
Here $F$ is the common (population-averaged) photon flux, and each $\mathcal{L}_{k}$ is evaluated by scanning $F$ while refitting nuisance parameters in the background model (diffuse normalizations and relevant nearby sources), with the photon index fixed to $-2$.

In the absence of a significant collective detection (we adopt ${\rm TS}_{\rm stack}<25$ as a non-detection criterion), we derive the 95\% C.L. UL $\hat{F}_{95}$ on the common flux from the stacked profile likelihood by using the same profile-likelihood method described in Section \ref{subsec:upperlimits}. We also produce stacked SEDs over 500~s intervals by performing the same likelihood scan independently in each energy bin across all stacked bursts (bin-by-bin SED). For the 500~s integration, we can stack four events (Figure \ref{fig:sed_500s}); GRB~200415A is excluded from the stacking because it is a known LAT detection and would dominate the combined constraint, whereas the remaining candidates yield only ULs in the GeV band. The two candidates excluded from the 500\,s stack (GRBs\,120616A and 200423A) lack LAT coverage within the first 500\,s post-trigger (Table~\ref{tab:lat_results}). 
\subsection{Photon Triplet Counting Analysis}\label{subsec:triplet}

We perform a model-independent \chng{photon-triplet} search designed to identify short-duration GeV emission that may be temporally coincident with MGF events. This approach complements the time-integrated likelihood analysis by providing sensitivity to brief, intense GeV flares that could be diluted in longer integration windows. The method was originally developed for the \fermi--LAT analysis of GRB~200415A \citep{Ajello:2021} and subsequently adapted for searches for gamma-ray counterparts to fast radio bursts \citep{Principe:2023}. The procedure is the following.

For each MGF candidate, we extract all \chng{SOURCE-class (\texttt{P8R3\_SOURCE})} LAT photons in the energy range 0.1--100~GeV within a circular region of interest (ROI) of radius $1^\circ$ centered on the host galaxy coordinates. We search for \chng{sets} of three consecutive photons (hereafter ``triplets'') in the selected region around the target.

For a given set of three consecutive photons indexed as $i$, $i+1$, and $i+2$, we define the triplet time interval as
\begin{equation}
\Delta t_i = t_{i+2} - t_i,
\end{equation}
which characterizes the \chng{waiting time of the triplet}. Under the assumption of a steady Poisson background with mean rate $R$ (estimated from the ROI and observation period), \chng{$\Delta t$ is the sum of two exponential waiting times and is therefore Erlang-2 distributed, with} probability density
\begin{equation}
P(\Delta t) = R^2 \, \Delta t \, e^{-R \Delta t},
\end{equation}
\chng{so that the probability of observing a triplet at least as compact as $\Delta t$ is
\begin{equation}
F(\Delta t) = 1 - e^{-R \Delta t}\,(1 + R \Delta t)
\end{equation}
\citep{LiMa:1983ApJ272317, Principe:2023, Xing:2024}.}

Following \citet{LiMa:1983ApJ272317}, we quantify the statistical significance of each triplet using the Li \& Ma test statistic, which accounts for both the observed photon excess and the estimated background rate in the ROI. The Li \& Ma significance is expressed in Gaussian-equivalent standard deviations ($\sigma$). 
\chng{Separately, we report the analytic waiting-time false-alarm probability $p = F(\Delta t)$ of the triplet defined above.}
\chng{The statistical power of the test resides in that a genuine GeV transient signal will have several photons within a short interval, 
yielding a $\Delta t$ far smaller than expected from the sparse background rate at these positions (see Figure \ref{fig:mgf_triplet_demographics}).}

A unique aspect of our MGF sample is that four events (GRB~081213A, GRB~180128A, GRB~200415A, and GRB~231024A) originate from the same host galaxy, NGC~253, corresponding to multiple flares from a single underlying magnetar. Because these events share the same sky position, the standard triplet search, which examines photons from a fixed spatial region, can suffer from temporal cross-contamination: photons from one flare may spuriously appear as appear as ``delayed'' triplets relative to an earlier trigger. \chng{In addition to the first-triplet search, we examine the most compact (minimum-$\Delta t$)
triplet within a one-year post-trigger window. Because this is a minimum
selected over the ${\sim}N_{\rm trip}$ triplets searched, it carries a look-elsewhere penalty, which
we evaluate directly by a photon-time scramble: redistributing the photon arrival times uniformly
over the livetime (corrected for bad time intervals), and recording the minimum triplet interval over
$10^{5}$ realizations yields a trials-corrected false-alarm probability
$p$ for each trigger window. The number of
realizations is set a priori so that the Monte-Carlo uncertainty on the
significance is $<0.02\sigma$ even for the smallest scramble-derived $p$ in the sample ($p=0.005$ for GRB\,231024A).
The photon-time scramble directly provides these per-window trials-corrected probabilities; the analytic Erlang-2 probability $F(\Delta t)$ is reported separately as the single-triplet, pre-trial compactness probability. Where noted,
we further apply a Bonferroni correction over the six non-detected events.}

\section{Results}\label{sec:results}

Table~\ref{tab:lat_results} summarizes the unbinned likelihood analysis results for each burst and integration window.

\textit{GRB\,200415A:} We detect a \lat\ source spatially consistent with the GBM localization in the longer integration windows. Across the $\Delta T=10^{2}$--$10^{3}$\,s windows, the TS values track the onset of the emission, culminating in a detection with $\mathrm{TS}\simeq25$ in the $10^3$\,s window. The measured photon flux is $F_\gamma=(3.6\pm2.0)\times10^{-6}\,\mathrm{ph\,cm^{-2}\,s^{-1}}$ with a hard photon index of $\Gamma=-1.71\pm0.34$. In the $\Delta T=10^{4}$\,s window, we continue to observe a significant excess ($\mathrm{TS}=22$), with $F_\gamma=(1.62\pm0.96)\times10^{-6}\,\mathrm{ph\,cm^{-2}\,s^{-1}}$ and $\Gamma=-1.67\pm0.36$. These properties are fully consistent with published \lat\ analyses of this event \citep{Ajello:2021,Chand:2021RAA,Zhang:2020ApJ}, which attribute the emission to synchrotron radiation from an external shock driven by relativistic ejecta or a shock resulting from collision with existing bowshock shell.

\textit{Other Candidates:} We find no evidence of GeV emission from GRBs\,231115A, 180128A, 120616A, 200423A, 231024A, or 081213A in any of the searched windows. Accordingly, we report ULs in Table~\ref{tab:lat_results}. For three events, GRBs 180128A, 120616A, and 200423A, the earliest emission is unconstrained because during the shorter integration windows we have no exposure after standard selections. The first \lat\ exposure for these events begins at $\sim1.5\times10^{2}$\,s, $\sim1.4\times10^{3}$\,s, and $\sim6.2\times10^{2}$\,s, respectively.
The limits reported here apply strictly to the observation times following the exposure. For these partial exposures, assuming no emission occurred during the unobserved interval, the reported limits are less constraining than those that would have been obtained from 
the full integration window.

\begin{figure}[t!] \centering \includegraphics[width=0.45\textwidth]{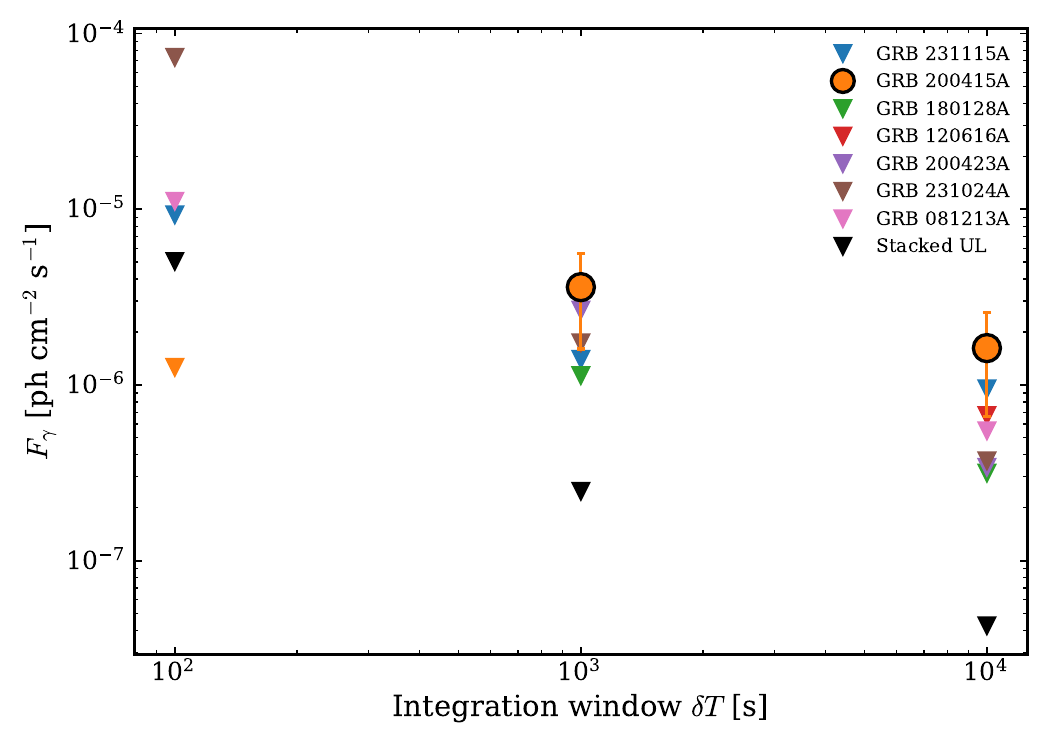}
\caption{
Photon fluxes (0.1--10\,GeV) from unbinned likelihood fits in post-trigger windows $\Delta T=10^{2},10^{3},10^{4}$~s. Filled circles show the GRB~200415A detections (1$\sigma$ statistical uncertainties). Colored inverted triangles show 95\% confidence ULs for the individual non-detected candidates (photon index fixed to $\Gamma=-2$); symbols with black outlines could be included in the stacked-likelihood analysis for the corresponding time window. Black inverted triangles show the 95\% confidence ULs from the stacked analysis. Note that at $\Delta T = 10^{2}\,$s only two events (GRBs\,081213A and 231115A) effectively contribute to the stacking likelihood; the remaining candidate (GRB\,231024A) has zero detected counts and non-zero exposure, so its individual UL -- derived from Poisson statistics on the exposure alone -- is inherently loose and does not meaningfully tighten the stacked constraint.
}
\label{fig:flux_vs_time}
\end{figure}

\begin{figure}[t!] \centering \includegraphics[width=0.45\textwidth]{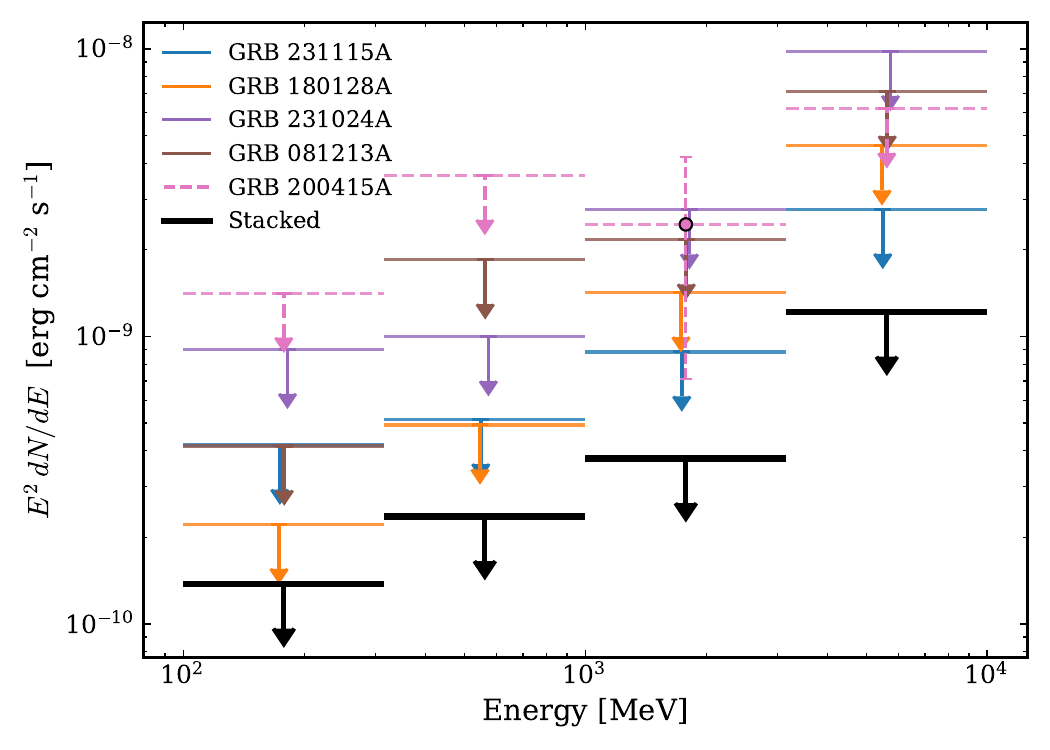}
\caption{
SED constraints in a 500\,s post-trigger interval. Downward arrows indicate 95\% C.L. ULs in each energy bin from the profile-likelihood scans with the index fixed to $\Gamma=-2$. The stacked limits combine the four non-detected events included in the 500\,s stack; GRB~200415A is shown for reference but excluded from the stacking.}
\label{fig:sed_500s}
\end{figure}

Figure~\ref{fig:flux_vs_time} displays the photon flux as a function of the integration time window $\delta T$ for the full sample. Circles denote the detection of GRB\,200415A, while inverted triangles represent ULs for the remaining candidates. 
The TS maps (generated with and without a point source at the host position) reveal no significant excesses beyond GRB\,200415A at the target positions (Figure~\ref{fig:tsmaps_500s}). The individual 500\,s SEDs also yield only ULs for all candidates except GRB\,200415A (Figure~\ref{fig:sed_500s}). To probe for emission below the individual detection thresholds, we consider the 
500\,s interval post-trigger and computed the stacked flux UL  and the energy-resolved SEDs (in black in Figures~\ref{fig:flux_vs_time}--\ref{fig:sed_500s}).

The complete triplet analysis results are presented in Table~\ref{tab:triplet_results}. Among the seven candidates, only GRB~200415A shows significant 
prompt GeV emission: \chng{its first post-trigger triplet has a waiting-time false-alarm probability $p = 3.8\times10^{-5}$ ($3.96\sigma$), with} a 
triplet time interval $\Delta t = 265$~s arriving within $\sim$20~s of the GBM
trigger\chng{; the same triplet gives $\sigma = 5.39$ under the Li\&Ma statistic evaluated for this triplet}. This provides \chng{a complementary
line of evidence for} the likelihood-based detection \citep{Ajello:2021}.

\begin{figure}[t!]
\centering
\includegraphics[width=\linewidth]{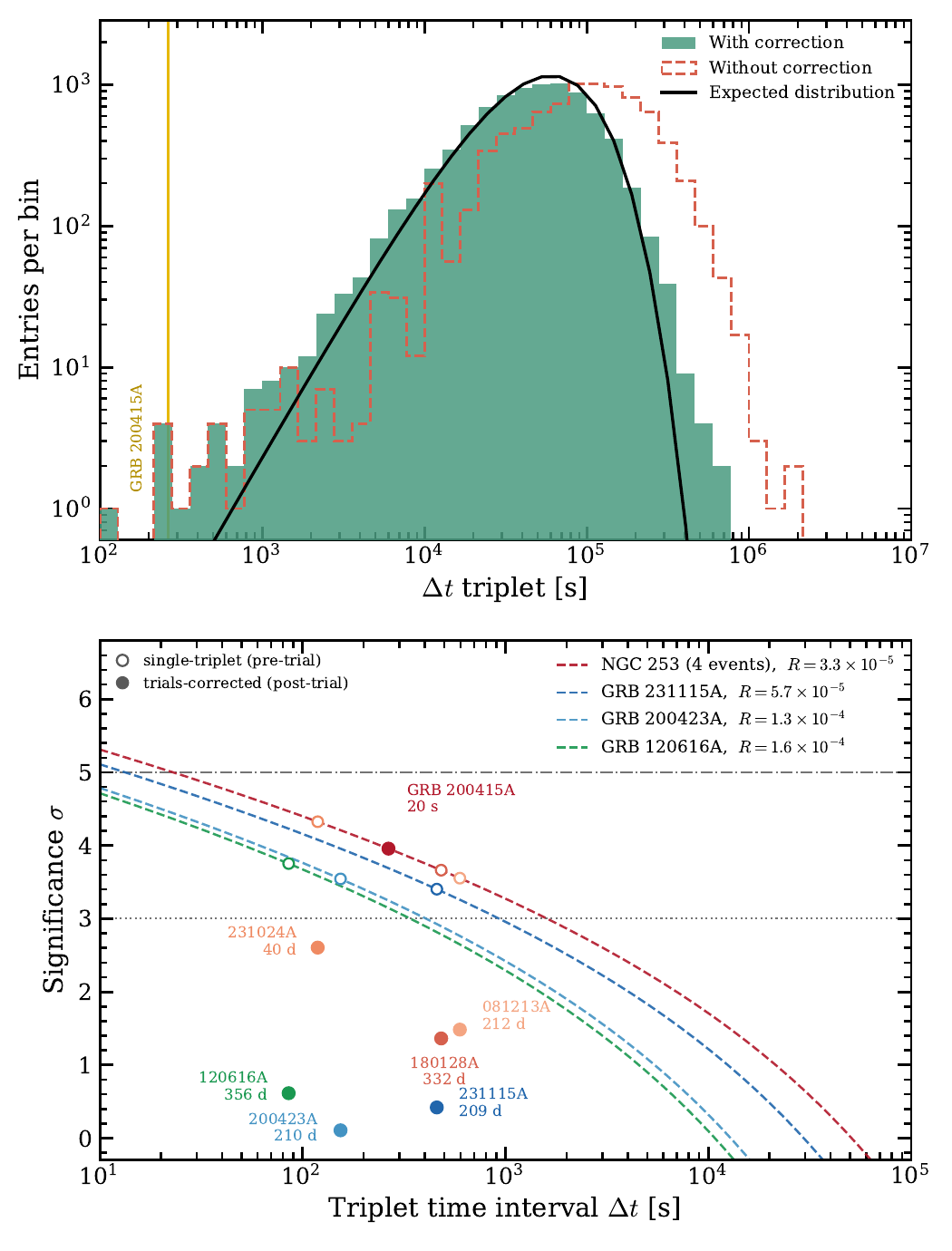}
\caption{\chng{Photon-triplet analysis at the MGF positions. \textit{Top:} distribution of the
triplet time interval $\Delta t$ for consecutive-photon triplets at NGC~253
(8375 triplets over 17\,yr), with (green) and without (dashed) the correction for \textit{Fermi}--LAT
orbit and field-of-view gaps, compared to the distribution expected for independent Poisson events
(solid black; Erlang-2). The correction shifts the distribution toward the expected form; the gold vertical line marks the GRB\,200415A first
post-trigger triplet ($\Delta t = 265$\,s). \textit{Bottom:} single-triplet waiting-time significance,
$\sigma(\Delta t) = \Phi^{-1}[1 - F(\Delta t)]$ with $F$ the Erlang-2 CDF, as a function of $\Delta t$
at each position's background rate $R$ (dashed curves; the four NGC~253 events share one rate).
Open circles mark the single-triplet (pre-trial) value on each curve; filled circles mark the 
per-window trials-corrected (post-trial) value from the one-year photon-time scramble. The
pre-specified first triplet of GRB\,200415A lies on its curve at $3.96\sigma$, 
whereas the six next-best triplets fall below their curves and are at delays of
40--356 days, consistent with background.}}
\label{fig:mgf_triplet_demographics}
\end{figure}

The remaining six candidates show no significant triplet emission, with $\sigma < 1$ and $\Delta t > 10^4$~s in all cases (Table~\ref{tab:triplet_results}). \chng{Examining instead the most compact (minimum-$\Delta t$) triplet within a window from the trigger, the most significant is GRB~231024A ($\Delta t = 119$~s, 40.5~days post-trigger), with a per-window trials-corrected false-alarm probability $p = 0.005$ ($2.6\sigma$) from the photon-time scramble; after a Bonferroni correction over the six non-detection trigger windows this becomes $p_{\rm global} = 0.03$ ($1.9\sigma$). All other windows reach at most $1.5\sigma$, consistent with background (Figure~\ref{fig:mgf_triplet_demographics}).} These results corroborate the likelihood-based non-detections.

\section{Discussion}\label{sec:discussion}

\textit{Phenomenological considerations}. Our LAT analysis yields a single GeV detection---GRB\,200415A---whose properties (Table~\ref{tab:lat_results}) agree with previous work. The $\sim$20\,s onset and emission extending over $\sim$20--400\,s are consistent with synchrotron radiation from an external shock or bow-shock driven by relativistic ejecta \citep{Zhang:2020ApJ,Ajello:2021,
Chand:2021RAA}. For the remaining candidates, the non-detections are consistent across all four search methods (fixed-window likelihood, 500\,s SED scans, stacking, and \chng{photon-triplet analysis}) reinforcing the absence of GeV emission from these events. We note, however, that for three of 
the six non-detected candidates 
(GRBs\,180128A, 120616A, and 200423A), the LAT 
exposure begins only at $\sim$148, $\sim$1447, and 
$\sim$619\,s post-trigger, respectively.

In the $10^{3}$\,s interval, the 0.1--10\,GeV 
energy-flux ULs are of order 
$10^{-9}\,\mathrm{erg\,cm^{-2}\,s^{-1}}$. In the 
$10^{4}$\,s window, the longer exposure yields deeper 
constraints: for GRB\,120616A in IC\,342 
($d\simeq2.3$\,Mpc) we obtain 
$F_E<5\times10^{-10}\,\mathrm{erg\,cm^{-2}\,s^{-1}}$, 
a factor of $\sim$4 below the $10^{4}$\,s flux of 
GRB\,200415A, while for GRBs\,081213A and 231024A in NGC\,253 ($d\simeq3.7$\,Mpc, the same host as 
GRB\,200415A) we find 
$F_E\lesssim 3\text{--}4\times10^{-10}\,
\mathrm{erg\,cm^{-2}\,s^{-1}}$. Since these events 
share the same distance as GRB\,200415A, the flux 
ULs translate directly into luminosity 
constraints, ruling out a GeV counterpart of 
comparable brightness. The stacked-likelihood analysis 
further strengthens this conclusion: combining the 
five candidates with LAT exposure in the $10^{3}$\,s 
window yields ${\rm TS}_{\rm stack}\approx 0$ and an 
UL
$\hat{F}_{95}\approx2\times10^{-10}\,
\mathrm{erg\,cm^{-2}\,s^{-1}}$ on the 
population-averaged 0.1--10\,GeV energy flux, 
implying that the UL on the average GeV emission across the
population is more than an order of magnitude below
that of GRB\,200415A. 

\textit{Physical interpretation}.

For each MGF in our sample, we use the observed prompt emission to estimate basic fireball parameters following \citet{Nakar:2005ApJ, Ioka:2005ApJ}. We take $E_{\gamma,\mathrm{iso}}$ from the prompt measurements reported in the literature and infer the characteristic prompt-spike luminosity as $L_{\gamma,\mathrm{iso}}\equiv E_{\gamma,\mathrm{iso}}/T_{90}=10^{47}L_{\gamma,47}\,\mathrm{erg\,s^{-1}}$ (Table~\ref{tab:mgf_sample_meta}). A fraction $\xi_{\gamma}$ of the total luminosity is assumed to emerge in $\gamma$ rays, so that the total fireball luminosity is
\(
L_0 = L_{\gamma,\mathrm{iso}}/\xi_{\gamma}
\simeq 3\times10^{47}\,
\xi_{\gamma,-0.5}^{-1}\,L_{\gamma,47}\,
\mathrm{erg\,s^{-1}},
\)
where we write $\xi_{\gamma}=0.3\,\xi_{\gamma,-0.5}$ as a fiducial normalisation (so that $1/\xi_\gamma = 1/0.3 \approx 3.3$, rounded to 3 in the prefactor).
The energy is initially released in a compact region of size $r_0 = 10^{6}\,r_{0,6}\,\mathrm{cm}$ (i.e., $r_{0,6}\equiv r_0/10^{6}\,\mathrm{cm}$) near the magnetar surface, producing a pair-rich, optically thick fireball. The initial (radiation-dominated) temperature is
\(
T_0 = \left( L_0/4\pi r_0^{2} c\, a \right)^{1/4}
\simeq 275\,
\xi_{\gamma,-0.5}^{-1/4}\,
L_{\gamma,47}^{1/4}\,
r_{0,6}^{-1/2}\ \mathrm{keV},
\)
where $a=7.56\times10^{-15}\,\mathrm{erg\,cm^{-3}\,K^{-4}}$ is the radiation constant. For the MGFs considered here, $T_0$ lies in the few 
$\times10^{2}$\,keV range 
(Table~\ref{tab:fireball_baryon}), and the observed 
$E_{\rm p}$ values are of the order of $3kT_0$ 
\citep{Zhang:2020ApJ}. The low-energy spectral indices 
of five candidates ($\alpha \approx -0.6$ to $+0.0$) 
are harder than the slow-cooling synchrotron limit 
($\alpha = -2/3$; \citealt{Preece:1998}), consistent 
with a quasi-thermal photospheric origin. The 
exceptions are GRBs\,200423A and 231024A, whose 
indices ($\alpha = -0.7$ and $-0.8$, respectively) 
fall below this limit, admitting a non-thermal origin 
for these two events.

As the fireball expands and accelerates, the Thomson optical depth decreases and the flow eventually becomes transparent. The critical Lorentz factor corresponding to the photospheric transition can be computed as
\(
\eta_*=\left(L_0 \sigma_{\rm T}/4\pi m_p c^{3} r_0\right)^{1/4}
\simeq 140\,
\xi_{\gamma,-0.5}^{-1/4}\,
L_{\gamma,47}^{1/4}\,
r_{0,6}^{-1/4},
\)
where $\sigma_{\rm T}$ is the Thomson cross section and $m_p$ is the proton mass, and the values we derive for each event is reported in Table~\ref{tab:fireball_baryon}. This critical value separates two limiting regimes for the outflow composition, parametrized by the dimensionless entropy $\eta \equiv L_0/\dot{M}c^2$. In particular, in the baryonic-poor regime ($\eta>\eta_*$)\footnote{This regime is also the one required to reproduce the prompt scaling $E_p\propto E_{\rm iso}^{1/4}$ reported for magnetar giant flares \citep{Zhang:2020ApJ, Trigg:2025}, which requires a photosphere located within the acceleration phase \citep{Zhang:2020ApJ}.} the photosphere lies in the acceleration phase and the ejecta kinetic energy is suppressed relative to the injected fireball energy. This implies that the fireball radiates away the majority of its internal energy before it can fully accelerate. The terminal Lorentz factor saturates at $\Gamma_{\rm ej} \simeq \eta_*$, and the fraction of the total luminosity converted into bulk kinetic energy is suppressed by the factor $\eta_*/\eta$ \citep{Ioka:2005ApJ}.  Under these conditions, the isotropic baryonic kinetic energy is therefore limited to \(E_{b,\mathrm{iso}} \simeq (\eta_*/\eta) E_0\), which allows to impose a strict physical UL on the baryonic mass $M_b$ carried by the ejecta. Since a higher baryonic load would transition the fireball into the baryonic-rich regime, the mass must satisfy \(M_b < M_b^{\rm (BP)} \equiv (E_{\gamma,\mathrm{iso}}/\xi_\gamma \eta_* c^2)\).
Using the prompt parameters from Table~\ref{tab:mgf_sample_meta} and the derived $\eta_*$ values in Table~\ref{tab:fireball_baryon}, we find that the mass limits are $M_{b,22}^{\rm (BP)} \lesssim 1.3$--$3.9$ (in units of $10^{22}$\,g) for the fainter flares and $\lesssim 40.8$ for the bright event GRB\,200415A. \chng{The inferred baryonic-poor parameters depend on the 
assumed prompt radiative efficiency $\xi_\gamma$ and initial fireball 
radius $r_0$ as $\eta_* \propto \xi_\gamma^{-1/4} r_0^{-1/4}$ and $M_b^{\rm (BP)} \propto \xi_\gamma^{-3/4} r_0^{1/4}$. Varying $r_0$ from $10$ to $15$~km 
changes $\eta_*$ and $M_b^{\rm (BP)}$ by only approximately $10\%$. Relative to the fiducial value $\xi_\gamma=0.3$, adopting $\xi_\gamma=0.1$, $0.5$, and $1$ 
changes the mass upper limits by factors of $2.28$, $0.68$, and $0.41$, respectively. Thus, over the representative parameter range considered here, the 
mass limits vary by at most a factor of approximately $2.3$ relative to the fiducial values and remain of the same order of magnitude.}

We now compare the physical baryonic-poor requirement ($\eta>\eta_*$) with the observational constraints set by LAT non-detections. The latter provide ULs on the 0.1--10\,GeV radiated energy and therefore on the shock-powered energy budget, which we express as limits on $M_b$ (and equivalently on $E_{\mathrm{k,iso}}$) for assumed efficiencies \citep{Nakar:2005ApJ, Ioka:2005ApJ}. Assuming that a fraction \(\xi_L\) of the baryonic kinetic energy is radiated in the GeV band via the shock-powered component, the observed LAT energy (or UL) implies a corresponding bound on the ejecta mass,
\( M_b < M_b^{\rm (LAT)} = E_{\mathrm{LAT,iso}}/(\xi_L \eta_* c^2) \).
We adopt a fiducial conversion efficiency of \(\xi_L = 0.1\), which renders the LAT-derived limits conservative. We convert the measured 0.1--10\,GeV energy fluxes (and ULs) into isotropic-equivalent energies using the host distances in Table~\ref{tab:mgf_sample_meta}. Adopting the same bow-shock environment as 
GRB\,200415A ($f \approx 50$), the five quasi-thermal 
candidates yield $t_{\rm coll} \sim 10$\,s, 
$\Gamma_{\rm sh} \sim 10$, $t_\theta \sim 500$\,s, 
and $E_{\rm syn,max} \sim 1$\,GeV. The predicted emission from the bow-shock interaction is anticipated within the first $\sim$1000\,s, and we therefore use our $10^3$\,s integration window for the LAT-derived constraints on the baryonic load. When the LAT-derived limit is weaker than the baryonic-poor upper bound (i.e., $M_b^{\rm (LAT)} \gg M_b^{\rm (BP)}$), the non-detection is expected: the physically allowed ejecta are intrinsically below current LAT sensitivity, and the observations do not probe the baryonic-poor constraint.

For the detected event GRB\,200415A, the observed GeV emission requires a minimum baryonic load to power the signal. We find $M_b^{\rm (LAT)} \gtrsim 5.8 \times 10^{23}$\,g (lower limit), which is comparable to the baryonic-poor UL $M_b^{\rm (BP)} \lesssim 4.1 \times 10^{23}$\,g. Within the systematic uncertainties of the efficiency parameters, the convergence of these two bounds implies that GRB\,200415A occurred in a marginally baryonic-poor regime where $\eta \sim \eta_*$. This parameter regime allows the fireball to be clean enough to exhibit high-$E_p$ prompt emission, yet sufficiently loaded to carry the kinetic energy required for a detectable GeV afterglow. A modest increase in the effective GeV efficiency (larger $\xi_L$) reconciles the apparent tension between $M_b^{\rm(LAT)}$ and $M_b^{\rm(BP)}$ while retaining $\eta\sim\eta_*$.

Conversely, for the non-detected events, the physical constraints diverge significantly. For a representative case like GRB\,081213A, the baryonic-poor physics restricts the mass to $M_{b,22}^{\rm (BP)} < 2.1$. However, the sensitivity of the LAT corresponds to a much higher observational threshold of $M_{b,22}^{\rm (LAT)} < 65.6$. Since the physically allowed mass is far below the sensitivity threshold ($M_b^{\rm (BP)} \ll M_b^{\rm (LAT)}$), the non-detections are a natural prediction of the model.
Even considering the deeper limits achieved via stacking analysis, the predicted signal from these clean outflows remains below the detection limit. This conclusion still holds if we apply moderate changes in the assumed spectral index used to derive the limits.  Thus, the lack of GeV detections for the broader population is consistent with the intrinsic faintness of baryonic-poor outflows and will remain undetectable with the current sensitivity. 

The two candidates whose spectral indices fall below 
the slow-cooling synchrotron limit---GRBs\,200423A 
($\alpha = -0.7$) and 231024A 
($\alpha = -0.8$)---may instead produce their prompt 
emission via optically thin processes such as 
synchrotron radiation from internal shocks or magnetic 
reconnection above the photosphere. We note that the 
typical uncertainties on $\alpha$ for these faint, 
short events ($\sigma_\alpha \sim 0.2$--$0.3$; 
\citealt{Trigg:2024, Trigg:2025}) place the 
slow-cooling boundary within $1\sigma$ of both central 
values, so a photospheric origin cannot be firmly 
excluded. Nevertheless, we consider the implications 
of a non-thermal interpretation. In this case, the 
photospheric scaling 
$E_{\rm p} \propto E_{\rm iso}^{1/4}$ does not apply, 
the baryonic-poor requirement $\eta > \eta_*$ is not 
enforced, and the outflow may reside in the 
baryonic-rich regime ($\eta < \eta_*$), where the ejecta 
reach $\Gamma_{\rm ej} \simeq \eta$ before becoming 
transparent and the kinetic energy is not suppressed, 
yielding 
$E_{\rm k,iso} \simeq E_{\gamma,\rm iso}
(1/\xi_\gamma - 1)$ 
(Table~\ref{tab:fireball_baryon}). Alternatively, if 
the prompt emission is powered by magnetic 
reconnection, a substantial fraction of the outflow 
energy is dissipated during the prompt phase, leaving 
a reduced energy budget for any subsequent afterglow. 
GRB\,200423A is the more energetic of the two 
($E_{\gamma,\rm iso} = 8.5\times10^{45}$\,erg); 
however, the LAT field of view did not include this 
source until $\sim$619\,s post-trigger, precluding 
meaningful constraints on early GeV emission. 
GRB\,231024A ($E_{\gamma,\rm iso} = 
5.5\times10^{44}$\,erg) had LAT exposure from 
$\sim$100\,s, yet yields $\mathrm{TS} = 0$ at all 
time windows (Table~\ref{tab:lat_results}), consistent 
with either a reconnection-dominated prompt phase in 
which the available energy has already been radiated, 
or an intrinsically faint afterglow reflecting its 
order-of-magnitude lower isotropic energy relative to 
GRB\,200415A.

\section{Conclusions}\label{sec:conclusions}

We carried out a systematic \textit{Fermi}--LAT search for high-energy ($0.1$--$10$\,GeV) emission from seven nearby extragalactic MGF candidates identified in
\textit{Fermi}--GBM data \citep{Trigg:2025}. For each target, we performed unbinned likelihood analyses in post-trigger integration windows of $10^{2}$, $10^{3}$, and $10^{4}$\,s, constructed standardized 500\,s data products including TS maps and energy-resolved SEDs, and applied both stacked-likelihood and model-independent photon-triplet \chng{searches}. 

Among the seven candidates, only GRB\,200415A shows significant GeV emission, with $\mathrm{TS}\simeq25$ in the $10^{3}$\,s window and a hard spectrum ($\Gamma \simeq -1.7$) consistent with previous LAT analyses and with synchrotron radiation from an external shock or bow-shock interaction \citep{Zhang:2020ApJ,Ajello:2021,Chand:2021RAA}. \chng{The photon-triplet analysis provides a complementary line
of evidence for this detection}, with the first triplet arriving $\sim$20\,s after the GBM trigger. For the remaining six candidates, no evidence of GeV emission is found in any searched interval or analysis method, with all triplet significances $\sigma < 1$, corroborating the likelihood-based non-detections. For individual events, we obtain 95\% C.L.\ energy-flux ULs of order $F_E\sim10^{-9}\,\mathrm{erg\,cm^{-2}\,s^{-1}}$ assuming a fiducial $\Gamma=-2$ spectrum. Stacking the four non-detected events with LAT exposure in the 500\,s interval, spanning distances $d\simeq 2.3$--$7.7$\,Mpc, yields a population-averaged UL of $\approx 2\times10^{-10}\,\mathrm{erg\,cm^{-2}\,s^{-1}}$, more than an order of magnitude below the measured flux of GRB\,200415A. We note that for three events (GRBs\,180128A, 120616A, and 200423A), the source position was outside the LAT field of view at trigger time; the reported limits apply only after the first LAT exposure and any earlier GeV emission remains unconstrained.

Within the relativistic fireball framework, the hard prompt spectra of five candidates favor baryonic-poor outflows ($\eta>\eta_{\ast}$), which restrict the ejecta mass to $M_b \lesssim 1$--$4\times10^{22}$\,g. This stringent baryon limit suppresses the kinetic energy available to drive a GeV afterglow well below current LAT sensitivity, naturally explaining the non-detections even with the deeper constraints achieved through stacking. GRB\,200415A is consistent with an outflow near the critical transition $\eta\sim\eta_{\ast}$, where the fireball is sufficiently clean to produce a high-$E_{\rm p}$ prompt spike yet carries enough baryonic load to power the observed GeV signal. Two candidates (GRBs\,200423A and 231024A) have softer spectral indices implying a non-thermal prompt origin and potentially a baryonic-rich regime; however, their non-detections are explained by either late LAT exposure onset or intrinsically low isotropic energy.

The results of this analysis suggest that, within the limited statistics of the present sample (one detection out of seven candidates), detectable GeV afterglows from MGFs may be restricted to the most energetic events with favorable observing conditions,
and that the lack of GeV counterparts in the broader population is a natural prediction of the baryonic-poor fireball model rather than an observational
limitation alone. However, for single events the number of GeV photons is very limited and the detection of such emission remains challenging with the current effective area of the LAT. As the extragalactic MGF sample continues to grow through ongoing GBM searches, future LAT observations and next-generation $\gamma$-ray instruments with larger effective areas may better address whether GRB\,200415A represents a rare extreme or the bright end of a broader population of GeV-emitting magnetar flares.

\begin{acknowledgments}
The \textit{Fermi} LAT Collaboration acknowledges generous ongoing support
from a number of agencies and institutes that have supported both the
development and the operation of the LAT as well as scientific data analysis.
These include the National Aeronautics and Space Administration and the
Department of Energy in the United States, the Commissariat \`a l'Energie Atomique
and the Centre National de la Recherche Scientifique / Institut National de Physique
Nucl\'eaire et de Physique des Particules in France, the Agenzia Spaziale Italiana
and the Istituto Nazionale di Fisica Nucleare in Italy, the Ministry of Education,
Culture, Sports, Science and Technology (MEXT), High Energy Accelerator Research
Organization (KEK) and Japan Aerospace Exploration Agency (JAXA) in Japan, and
the K.~A.~Wallenberg Foundation, the Swedish Research Council and the
Swedish National Space Board in Sweden.

Additional support for science analysis during the operations phase is gratefully
acknowledged from the Istituto Nazionale di Astrofisica in Italy and the Centre
National d'\'Etudes Spatiales in France. This work performed in part under DOE
Contract DE-AC02-76SF00515.
This work made use of public \textit{Fermi} data and the \textit{Fermi} Science Support Center (FSSC) analysis tools.
\end{acknowledgments}

\clearpage
\appendix
\renewcommand{\thefigure}{A\arabic{figure}}
\setcounter{figure}{0}
\section{Supplementary Figures}\label{sec:appendix}
\begin{figure*}[p]
    \centering
    \includegraphics[width=0.70\textwidth]{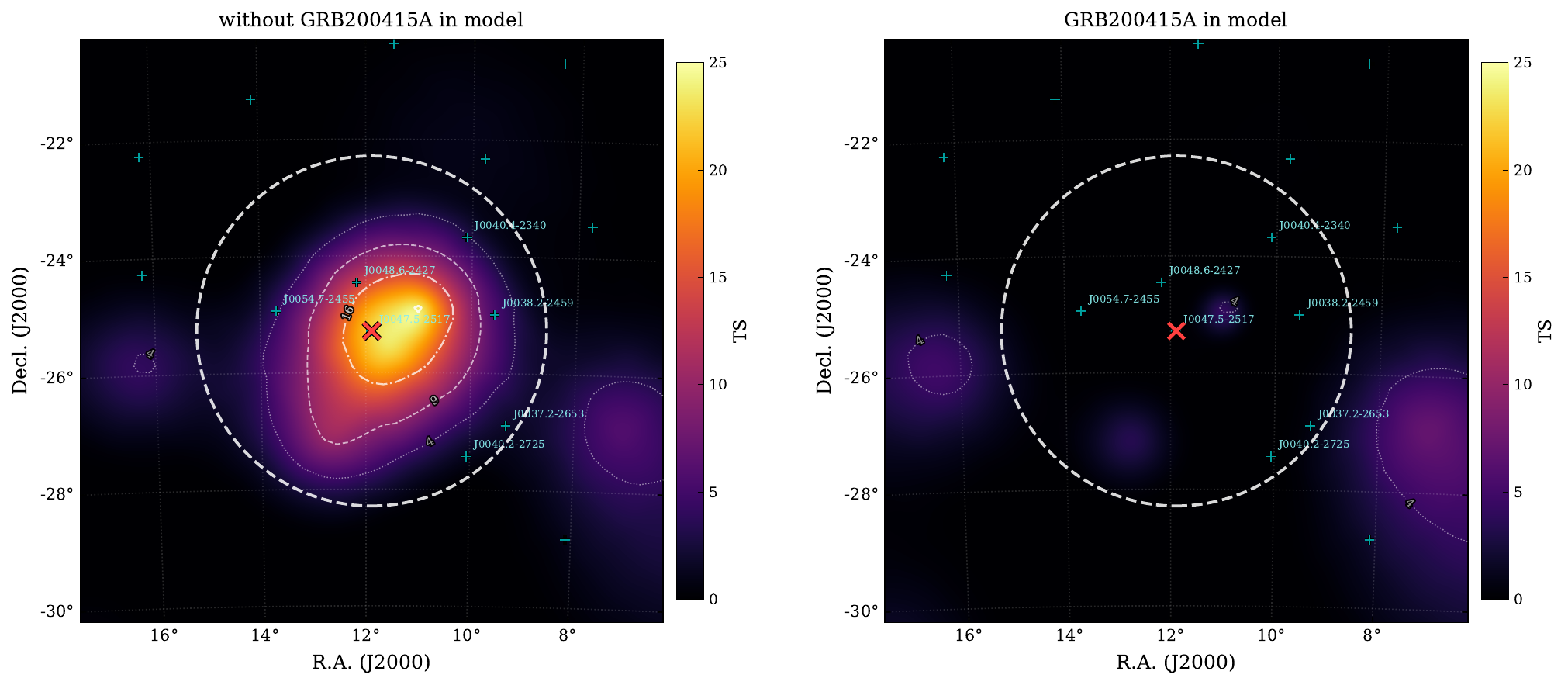}\\[4pt]\rule{0.7\textwidth}{0.4pt}\\[4pt]
    \includegraphics[width=0.32\textwidth]{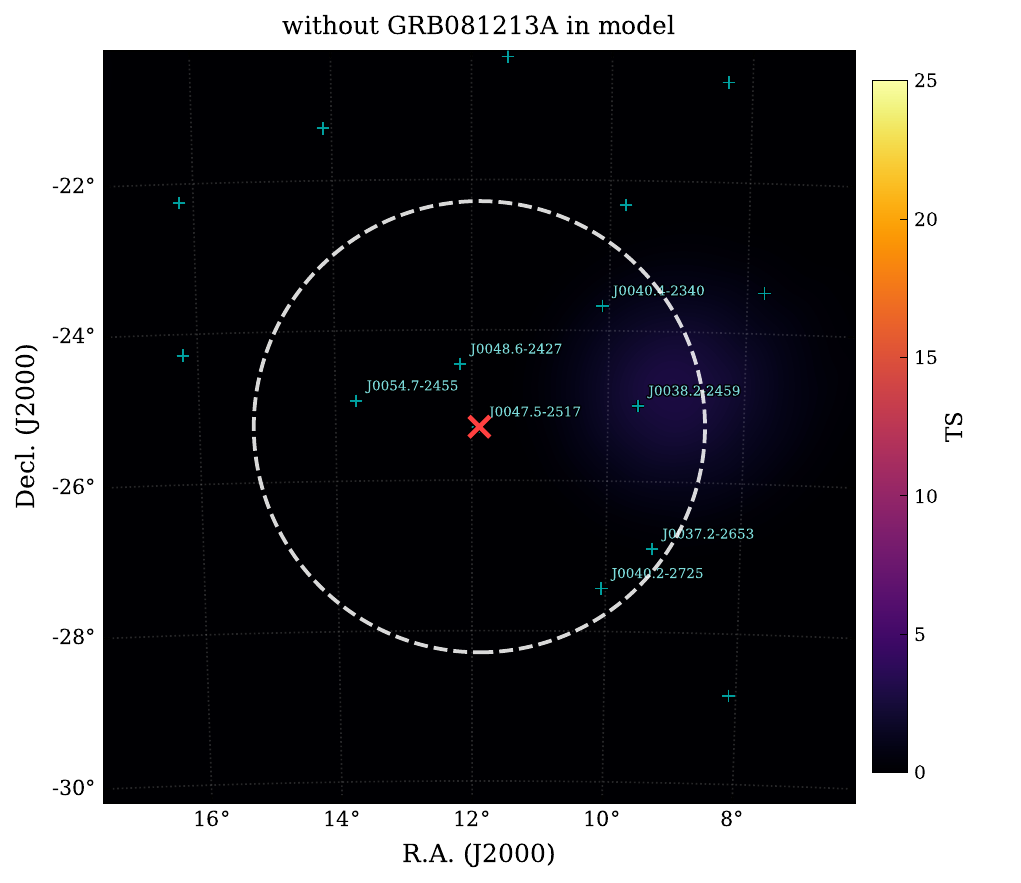}\hfill
    \includegraphics[width=0.32\textwidth]{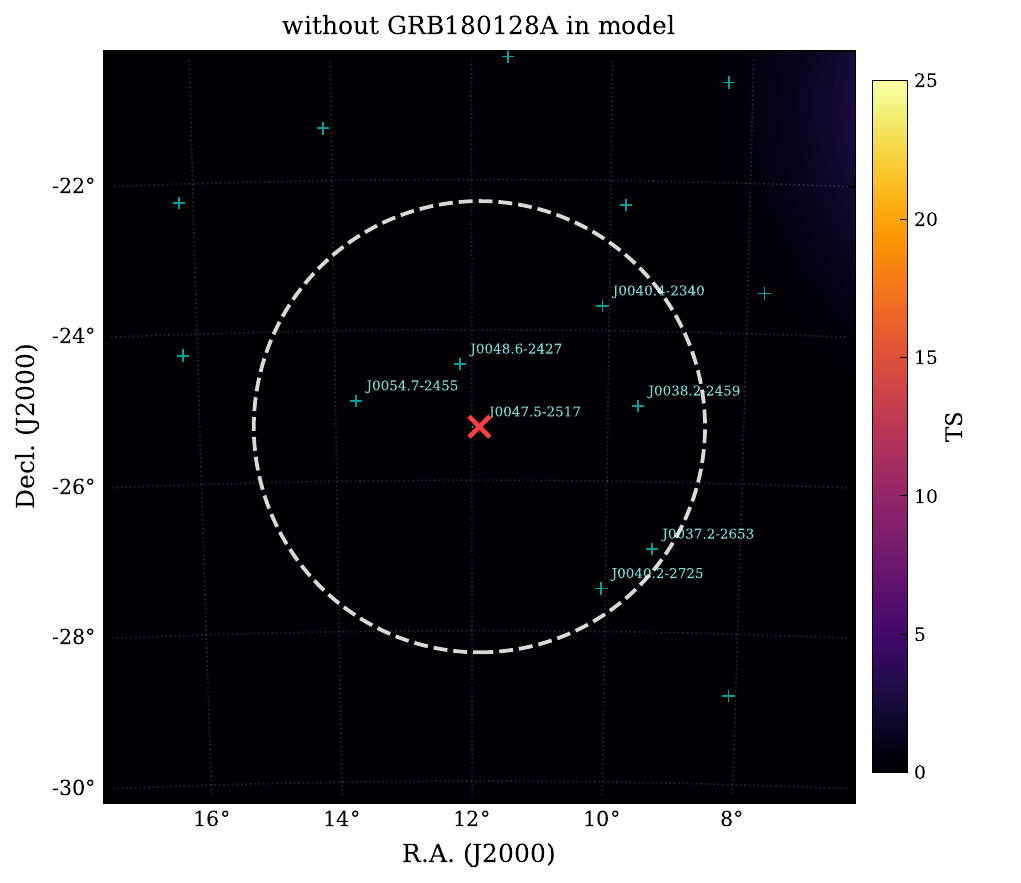}\hfill
    \includegraphics[width=0.32\textwidth]{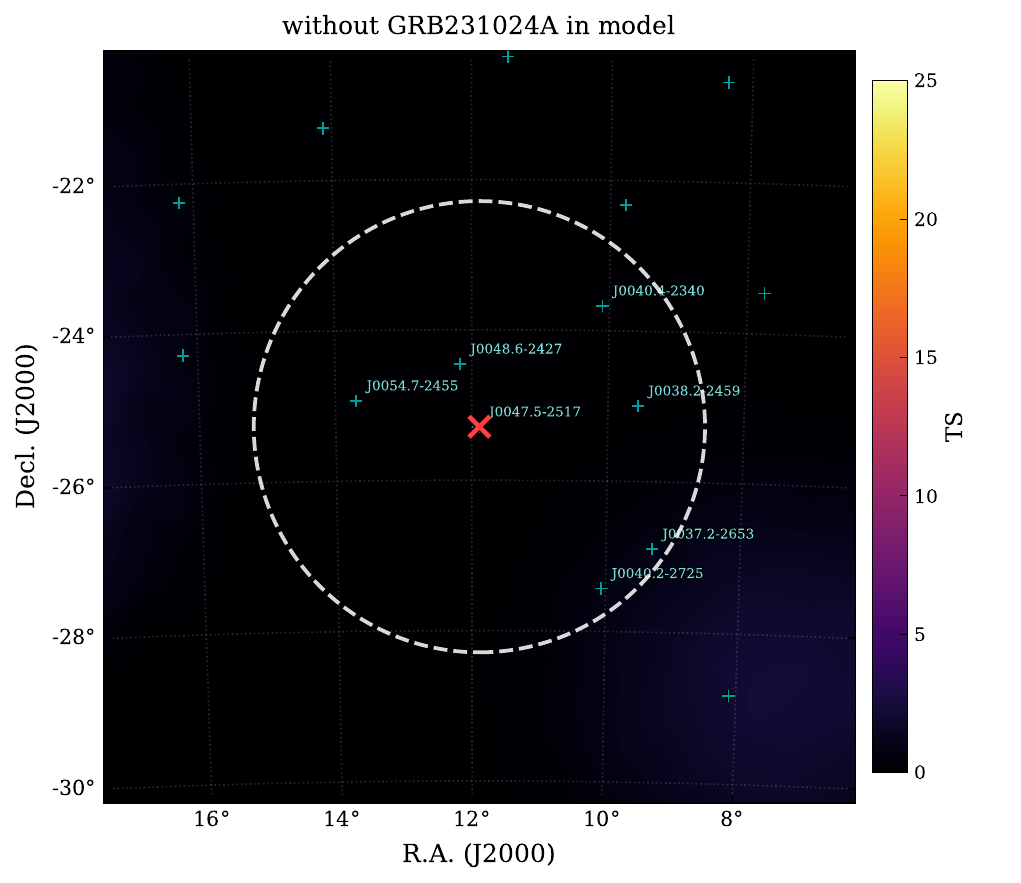}\\[4pt]
    \includegraphics[width=0.32\textwidth]{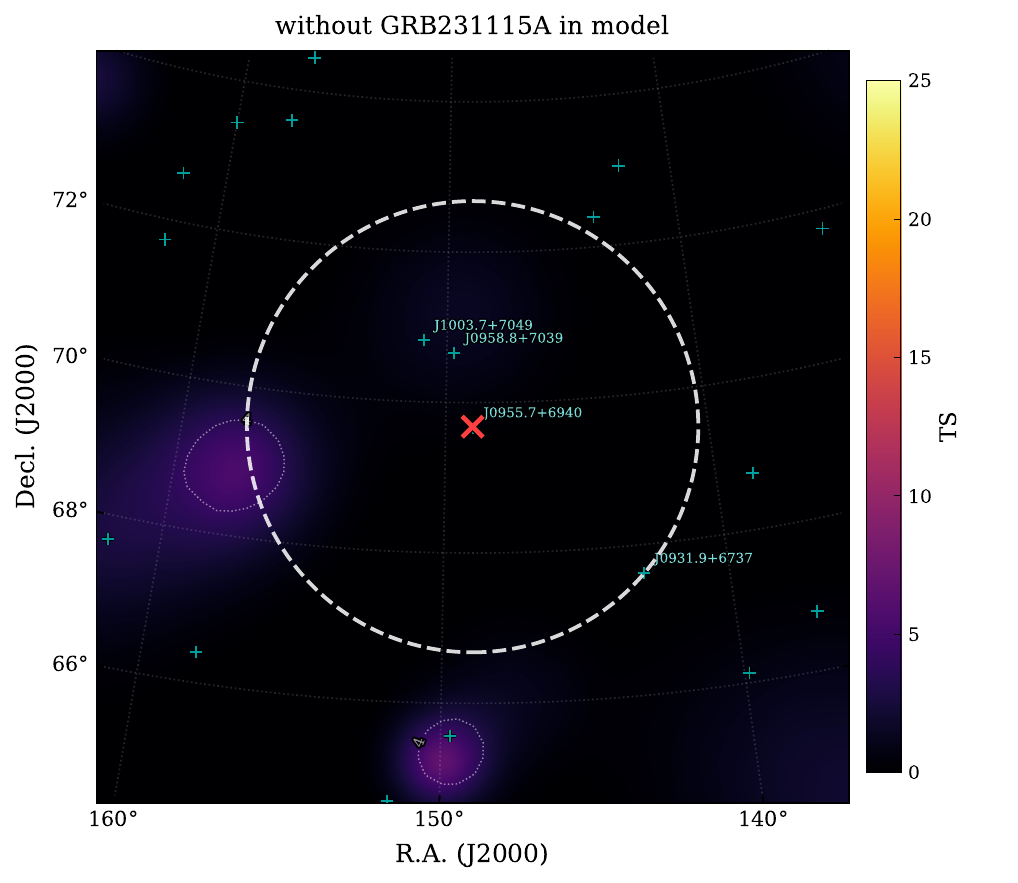}\hfill
    \includegraphics[width=0.32\textwidth]{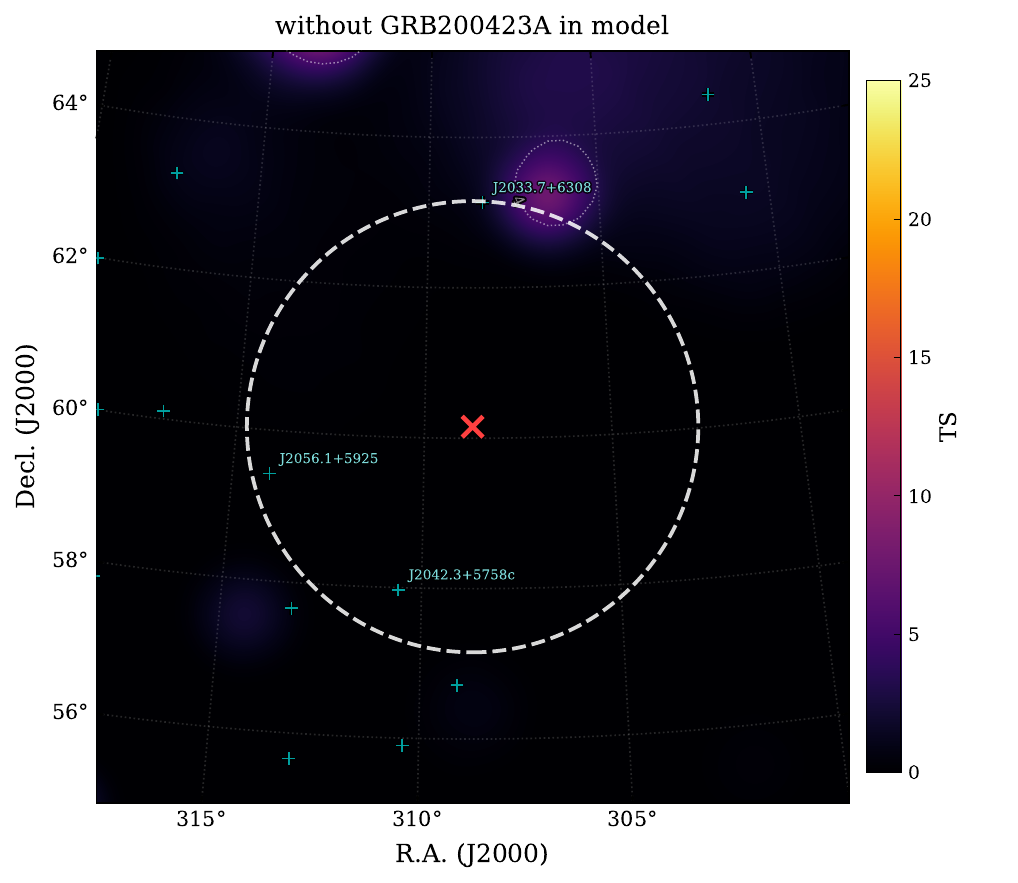}\hfill
    \makebox[0.32\textwidth]{}
\caption{\chng{LAT TS maps in the post-trigger window for each MGF candidate. \textit{Top:} the single
detection, GRB\,200415A, shown both without (left) and with (right) a transient point source at the
host position; the with-source map illustrates that the fitted source accounts for the excess.
\textit{Middle and bottom:} the five non-detections (GRB\,081213A, GRB\,180128A, GRB\,231024A;
GRB\,231115A, GRB\,200423A), for which only the without-source map is shown, indicating the
significance any excess would have to reach. Contours indicate TS levels of 4, 9, 16, and 25; the
$3^\circ$-radius circle is centered on the host coordinates. The window is $T_0$ to $T_0{+}500$\,s
for all events except GRB\,200423A, whose first LAT exposure begins at ${\sim}619$\,s and which
therefore uses $T_0{+}619$ to $T_0{+}1000$\,s.}}
    \label{fig:tsmaps_500s}
\end{figure*}

\bibliographystyle{aasjournal}
\bibliography{MGF_ms_v1}

\begin{thebibliography}{}
\expandafter\ifx\csname natexlab\endcsname\relax\def\natexlab#1{#1}\fi
\providecommand{\url}[1]{\href{#1}{#1}}
\providecommand{\dodoi}[1]{doi:~\href{http://doi.org/#1}{\nolinkurl{#1}}}
\providecommand{\doeprint}[1]{\href{http://ascl.net/#1}{\nolinkurl{http://ascl.net/#1}}}
\providecommand{\doarXiv}[1]{\href{https://arxiv.org/abs/#1}{\nolinkurl{https://arxiv.org/abs/#1}}}

\bibitem[{{Abdollahi} {et~al.}(2022){Abdollahi}, {Acero}, {Baldini}, {Ballet},
  {Bastieri}, {Bellazzini}, {Berenji}, {Berretta}, {Bissaldi}, {Blandford},
  {Bloom}, {Bonino}, {Brill}, {Britto}, {Bruel}, {Burnett}, {Buson}, {Cameron},
  {Caputo}, {Caraveo}, {Castro}, {Chaty}, {Cheung}, {Chiaro}, {Cibrario},
  {Ciprini}, {Coronado-Bl{\'a}zquez}, {Crnogorcevic}, {Cutini}, {D'Ammando},
  {De Gaetano}, {Digel}, {Di Lalla}, {Dirirsa}, {Di Venere}, {Dom{\'\i}nguez},
  {Fallah Ramazani}, {Fegan}, {Ferrara}, {Fiori}, {Fleischhack}, {Franckowiak},
  {Fukazawa}, {Funk}, {Fusco}, {Galanti}, {Gammaldi}, {Gargano}, {Garrappa},
  {Gasparrini}, {Giacchino}, {Giglietto}, {Giordano}, {Giroletti}, {Glanzman},
  {Green}, {Grenier}, {Grondin}, {Guillemot}, {Guiriec}, {Gustafsson},
  {Harding}, {Hays}, {Hewitt}, {Horan}, {Hou}, {J{\'o}hannesson}, {Karwin},
  {Kayanoki}, {Kerr}, {Kuss}, {Landriu}, {Larsson}, {Latronico},
  {Lemoine-Goumard}, {Li}, {Liodakis}, {Longo}, {Loparco}, {Lott}, {Lubrano},
  {Maldera}, {Malyshev}, {Manfreda}, {Mart{\'\i}-Devesa}, {Mazziotta}, {Mereu},
  {Meyer}, {Michelson}, {Mirabal}, {Mitthumsiri}, {Mizuno}, {Moiseev},
  {Monzani}, {Morselli}, {Moskalenko}, {Negro}, {Nuss}, {Omodei}, {Orienti},
  {Orlando}, {Paneque}, {Pei}, {Perkins}, {Persic}, {Pesce-Rollins},
  {Petrosian}, {Pillera}, {Poon}, {Porter}, {Principe}, {Rain{\`o}}, {Rando},
  {Rani}, {Razzano}, {Razzaque}, {Reimer}, {Reimer}, {Reposeur},
  {S{\'a}nchez-Conde}, {Saz Parkinson}, {Scotton}, {Serini}, {Sgr{\`o}},
  {Siskind}, {Smith}, {Spandre}, {Spinelli}, {Sueoka}, {Suson}, {Tajima},
  {Tak}, {Thayer}, {Thompson}, {Torres}, {Troja}, {Valverde}, {Wood}, \&
  {Zaharijas}}]{Abdollahi2022_4FGLDR3}
{Abdollahi}, S., {Acero}, F., {Baldini}, L., {et~al.} 2022, \apjs, 260, 53,
  \dodoi{10.3847/1538-4365/ac6751}

\bibitem[{{Ajello} {et~al.}(2019){Ajello}, {Arimoto}, {Axelsson}, {Baldini},
  {Barbiellini}, {Bastieri}, {Bellazzini}, {Bhat}, {Bissaldi}, {Blandford},
  {Bonino}, {Bonnell}, {Bottacini}, {Bregeon}, {Bruel}, {Buehler}, {Cameron},
  {Caputo}, {Caraveo}, {Cavazzuti}, {Chen}, {Cheung}, {Chiaro}, {Ciprini},
  {Costantin}, {Crnogorcevic}, {Cutini}, {Dainotti}, {D'Ammando}, {de la Torre
  Luque}, {de Palma}, {Desai}, {Desiante}, {Di Lalla}, {Di Venere}, {Fana
  Dirirsa}, {Fegan}, {Franckowiak}, {Fukazawa}, {Funk}, {Fusco}, {Gargano},
  {Gasparrini}, {Giglietto}, {Giordano}, {Giroletti}, {Green}, {Grenier},
  {Grove}, {Guiriec}, {Hays}, {Hewitt}, {Horan}, {J{\'o}hannesson}, {Kocevski},
  {Kuss}, {Latronico}, {Li}, {Longo}, {Loparco}, {Lovellette}, {Lubrano},
  {Maldera}, {Manfreda}, {Mart{\'\i}-Devesa}, {Mazziotta}, {Mereu}, {Meyer},
  {Michelson}, {Mirabal}, {Mitthumsiri}, {Mizuno}, {Monzani}, {Moretti},
  {Morselli}, {Moskalenko}, {Negro}, {Nuss}, {Ohno}, {Omodei}, {Orienti},
  {Orlando}, {Palatiello}, {Paliya}, {Paneque}, {Persic}, {Pesce-Rollins},
  {Petrosian}, {Piron}, {Poolakkil}, {Poon}, {Porter}, {Principe}, {Racusin},
  {Rain{\`o}}, {Rando}, {Razzano}, {Razzaque}, {Reimer}, {Reimer}, {Reposeur},
  {Ryde}, {Serini}, {Sgr{\`o}}, {Siskind}, {Sonbas}, {Spandre}, {Spinelli},
  {Suson}, {Tajima}, {Takahashi}, {Tak}, {Thayer}, {Torres}, {Troja},
  {Valverde}, {Veres}, {Vianello}, {von Kienlin}, {Wood}, {Yassine}, {Zhu}, \&
  {Zimmer}}]{Ajello2019LATGRBCat2}
{Ajello}, M., {Arimoto}, M., {Axelsson}, M., {et~al.} 2019, \apj, 878, 52,
  \dodoi{10.3847/1538-4357/ab1d4e}

\bibitem[{{Atwood} {et~al.}(2013){Atwood}, {Albert}, {Baldini}, {Tinivella},
  {Bregeon}, {Pesce-Rollins}, {Sgr{\`o}}, {Bruel}, {Charles}, {Drlica-Wagner},
  {Franckowiak}, {Jogler}, {Rochester}, {Usher}, {Wood}, {Cohen-Tanugi}, \&
  {Zimmer}}]{Atwood2013P8}
{Atwood}, W., {Albert}, A., {Baldini}, L., {et~al.} 2013, arXiv e-prints,
  arXiv:1303.3514, \dodoi{10.48550/arXiv.1303.3514}

\bibitem[{{Atwood} {et~al.}(2009){Atwood}, {Abdo}, {Ackermann}, {Althouse},
  {Anderson}, {Axelsson}, {Baldini}, {Ballet}, {Band}, {Barbiellini},
  {Bartelt}, {Bastieri}, {Baughman}, {Bechtol}, {B{\'e}d{\'e}r{\`e}de},
  {Bellardi}, {Bellazzini}, {Berenji}, {Bignami}, {Bisello}, {Bissaldi},
  {Blandford}, {Bloom}, {Bogart}, {Bonamente}, {Bonnell}, {Borgland},
  {Bouvier}, {Bregeon}, {Brez}, {Brigida}, {Bruel}, {Burnett}, {Busetto},
  {Caliandro}, {Cameron}, {Caraveo}, {Carius}, {Carlson}, {Casandjian},
  {Cavazzuti}, {Ceccanti}, {Cecchi}, {Charles}, {Chekhtman}, {Cheung},
  {Chiang}, {Chipaux}, {Cillis}, {Ciprini}, {Claus}, {Cohen-Tanugi},
  {Condamoor}, {Conrad}, {Corbet}, {Corucci}, {Costamante}, {Cutini}, {Davis},
  {Decotigny}, {DeKlotz}, {Dermer}, {de Angelis}, {Digel}, {do Couto e Silva},
  {Drell}, {Dubois}, {Dumora}, {Edmonds}, {Fabiani}, {Farnier}, {Favuzzi},
  {Flath}, {Fleury}, {Focke}, {Funk}, {Fusco}, {Gargano}, {Gasparrini},
  {Gehrels}, {Gentit}, {Germani}, {Giebels}, {Giglietto}, {Giommi}, {Giordano},
  {Glanzman}, {Godfrey}, {Grenier}, {Grondin}, {Grove}, {Guillemot}, {Guiriec},
  {Haller}, {Harding}, {Hart}, {Hays}, {Healey}, {Hirayama}, {Hjalmarsdotter},
  {Horn}, {Hughes}, {J{\'o}hannesson}, {Johansson}, {Johnson}, {Johnson},
  {Johnson}, {Johnson}, {Kamae}, {Katagiri}, {Kataoka}, {Kavelaars}, {Kawai},
  {Kelly}, {Kerr}, {Klamra}, {Kn{\"o}dlseder}, {Kocian}, {Komin}, {Kuehn},
  {Kuss}, {Landriu}, {Latronico}, {Lee}, {Lee}, {Lemoine-Goumard}, {Lionetto},
  {Longo}, {Loparco}, {Lott}, {Lovellette}, {Lubrano}, {Madejski}, {Makeev},
  {Marangelli}, {Massai}, {Mazziotta}, {McEnery}, {Menon}, {Meurer},
  {Michelson}, {Minuti}, {Mirizzi}, {Mitthumsiri}, {Mizuno}, {Moiseev},
  {Monte}, {Monzani}, {Moretti}, {Morselli}, {Moskalenko}, {Murgia},
  {Nakamori}, {Nishino}, {Nolan}, {Norris}, {Nuss}, {Ohno}, {Ohsugi}, {Omodei},
  {Orlando}, {Ormes}, {Paccagnella}, {Paneque}, {Panetta}, {Parent}, {Pearce},
  {Pepe}, {Perazzo}, {Pesce-Rollins}, {Picozza}, {Pieri}, {Pinchera}, {Piron},
  {Porter}, {Poupard}, {Rain{\`o}}, {Rando}, {Rapposelli}, {Razzano}, {Reimer},
  {Reimer}, {Reposeur}, {Reyes}, {Ritz}, {Rochester}, {Rodriguez}, {Romani},
  {Roth}, {Russell}, {Ryde}, {Sabatini}, {Sadrozinski}, {Sanchez}, {Sander},
  {Sapozhnikov}, {Parkinson}, {Scargle}, {Schalk}, \&
  {Scolieri}}]{Atwood:2009ApJ}
{Atwood}, W.~B., {Abdo}, A.~A., {Ackermann}, M., {et~al.} 2009, \apj, 697,
  1071, \dodoi{10.1088/0004-637X/697/2/1071}

\bibitem[{{Ballet} {et~al.}(2023){Ballet}, {Bruel}, {Burnett}, {Lott}, \& {The
  Fermi-LAT collaboration}}]{Ballet2023_4FGLDR4}
{Ballet}, J., {Bruel}, P., {Burnett}, T.~H., {Lott}, B., \& {The Fermi-LAT
  collaboration}. 2023, arXiv e-prints, arXiv:2307.12546,
  \dodoi{10.48550/arXiv.2307.12546}

\bibitem[{{Bissaldi} \& {von Kienlin}(2008)}]{Bissaldi:2008GCN}
{Bissaldi}, E., \& {von Kienlin}, A. 2008, GRB Coordinates Network, 8668, 1

\bibitem[{{Bruel} {et~al.}(2018){Bruel}, {Burnett}, {Digel}, {Johannesson},
  {Omodei}, \& {Wood}}]{Bruel2018P8}
{Bruel}, P., {Burnett}, T.~H., {Digel}, S.~W., {et~al.} 2018, arXiv e-prints,
  arXiv:1810.11394, \dodoi{10.48550/arXiv.1810.11394}

\bibitem[{{Chand} {et~al.}(2021){Chand}, {Joshi}, {Gupta}, {Yang}, {Dimple},
  {Sharma}, {Yang}, {Chakraborty}, {Zou}, {Shao}, {Yang}, {Zhang}, {Pandey},
  {Banerjee}, \& {Moneer}}]{Chand:2021RAA}
{Chand}, V., {Joshi}, J.~C., {Gupta}, R., {et~al.} 2021, Research in Astronomy
  and Astrophysics, 21, 236, \dodoi{10.1088/1674-4527/21/9/236}

\bibitem[{{Fermi-LAT Collaboration} {et~al.}(2021){Fermi-LAT Collaboration},
  {Ajello}, {Atwood}, {Axelsson}, {Baldini}, {Barbiellini}, {Baring},
  {Bastieri}, {Bellazzini}, {Berretta}, {Bissaldi}, {Blandford}, {Bonino},
  {Bregeon}, {Bruel}, {Buehler}, {Burns}, {Buson}, {Cameron}, {Caraveo},
  {Cavazzuti}, {Chen}, {Cheung}, {Chiaro}, {Ciprini}, {Costantin},
  {Crnogorcevic}, {Cutini}, {D'Ammando}, {de la Torre Luque}, {de Palma},
  {Digel}, {Di Lalla}, {Di Venere}, {Dirirsa}, {Fukazawa}, {Funk}, {Fusco},
  {Gargano}, {Giglietto}, {Gill}, {Giordano}, {Giroletti}, {Granot}, {Green},
  {Grenier}, {Griffin}, {Guiriec}, {Hays}, {Horan}, {J{\'o}hannesson}, {Kerr},
  {Kova{\v{c}}evi{\'c}}, {Kuss}, {Larsson}, {Latronico}, {Li}, {Longo},
  {Loparco}, {Lovellette}, {Lubrano}, {Maldera}, {Manfreda},
  {Mart{\'\i}-Devesa}, {Mazziotta}, {McEnery}, {Mereu}, {Michelson}, {Mizuno},
  {Monzani}, {Morselli}, {Moskalenko}, {Negro}, {Omodei}, {Orienti}, {Orlando},
  {Paliya}, {Paneque}, {Pei}, {Pesce-Rollins}, {Piron}, {Poon}, {Porter},
  {Principe}, {Racusin}, {Rain{\`o}}, {Rando}, {Rani}, {Razzaque}, {Reimer},
  {Reimer}, {Parkinson}, {Scargle}, {Scotton}, {Serini}, {Sgr{\`o}}, {Siskind},
  {Spandre}, {Spinelli}, {Tajima}, {Takahashi}, {Tak}, {Torres}, {Tosti},
  {Troja}, {Wadiasingh}, {Wood}, {Yassine}, {Yusafzai}, \&
  {Zaharijas}}]{Ajello:2021}
{Fermi-LAT Collaboration}, {Ajello}, M., {Atwood}, W.~B., {et~al.} 2021, Nature
  Astronomy, 5, 385, \dodoi{10.1038/s41550-020-01287-8}

\bibitem[{{Fermi Science Support Development Team}(2019)}]{fermitools}
{Fermi Science Support Development Team}. 2019, {Fermitools: Fermi Science
  Tools}, Astrophysics Source Code Library, record ascl:1905.011.
\newblock \doeprint{1905.011}

\bibitem[{{Gaensler} {et~al.}(2005){Gaensler}, {Kouveliotou}, {Gelfand},
  {Taylor}, {Eichler}, {Wijers}, {Granot}, {Ramirez-Ruiz}, {Lyubarsky},
  {Hunstead}, {Campbell-Wilson}, {van der Horst}, {McLaughlin}, {Fender},
  {Garrett}, {Newton-McGee}, {Palmer}, {Gehrels}, \&
  {Woods}}]{Gaensler:2005Natur}
{Gaensler}, B.~M., {Kouveliotou}, C., {Gelfand}, J.~D., {et~al.} 2005, \nat,
  434, 1104, \dodoi{10.1038/nature03498}

\bibitem[{{Gelfand} {et~al.}(2005){Gelfand}, {Lyubarsky}, {Eichler},
  {Gaensler}, {Taylor}, {Granot}, {Newton-McGee}, {Ramirez-Ruiz},
  {Kouveliotou}, \& {Wijers}}]{Gelfand:2005ApJ}
{Gelfand}, J.~D., {Lyubarsky}, Y.~E., {Eichler}, D., {et~al.} 2005, \apjl, 634,
  L89, \dodoi{10.1086/498643}

\bibitem[{{Granot} {et~al.}(2006){Granot}, {Ramirez-Ruiz}, {Taylor}, {Eichler},
  {Lyubarsky}, {Wijers}, {Gaensler}, {Gelfand}, \&
  {Kouveliotou}}]{Granot:2006ApJ}
{Granot}, J., {Ramirez-Ruiz}, E., {Taylor}, G.~B., {et~al.} 2006, \apj, 638,
  391, \dodoi{10.1086/497680}

\bibitem[{{Hurley} {et~al.}(1999){Hurley}, {Cline}, {Mazets}, {Barthelmy},
  {Butterworth}, {Marshall}, {Palmer}, {Aptekar}, {Golenetskii}, {Il'Inskii},
  {Frederiks}, {McTiernan}, {Gold}, \& {Trombka}}]{Hurley:1999Natur}
{Hurley}, K., {Cline}, T., {Mazets}, E., {et~al.} 1999, \nat, 397, 41,
  \dodoi{10.1038/16199}

\bibitem[{{Ioka} {et~al.}(2005){Ioka}, {Razzaque}, {Kobayashi}, \&
  {M{\'e}sz{\'a}ros}}]{Ioka:2005ApJ}
{Ioka}, K., {Razzaque}, S., {Kobayashi}, S., \& {M{\'e}sz{\'a}ros}, P. 2005,
  \apj, 633, 1013, \dodoi{10.1086/466514}

\bibitem[{{Kaspi} \& {Beloborodov}(2017)}]{Kaspi:2017}
{Kaspi}, V.~M., \& {Beloborodov}, A.~M. 2017, \araa, 55, 261,
  \dodoi{10.1146/annurev-astro-081915-023329}

\bibitem[{{Li} \& {Ma}(1983)}]{LiMa:1983ApJ272317}
{Li}, T.-P., \& {Ma}, Y.-Q. 1983, \apj, 272, 317, \dodoi{10.1086/161295}

\bibitem[{{Lyutikov}(2006)}]{Lyutikov:2006MNRAS}
{Lyutikov}, M. 2006, \mnras, 367, 1594,
  \dodoi{10.1111/j.1365-2966.2006.10069.x}

\bibitem[{{Mattox} {et~al.}(1996){Mattox}, {Bertsch}, {Chiang}, {Dingus},
  {Digel}, {Esposito}, {Fierro}, {Hartman}, {Hunter}, {Kanbach}, {Kniffen},
  {Lin}, {Macomb}, {Mayer-Hasselwander}, {Michelson}, {von Montigny},
  {Mukherjee}, {Nolan}, {Ramanamurthy}, {Schneid}, {Sreekumar}, {Thompson}, \&
  {Willis}}]{Mattox:1996}
{Mattox}, J.~R., {Bertsch}, D.~L., {Chiang}, J., {et~al.} 1996, \apj, 461, 396,
  \dodoi{10.1086/177068}

\bibitem[{{Mazets} {et~al.}(1979){Mazets}, {Golentskii}, {Ilinskii}, {Aptekar},
  \& {Guryan}}]{Mazets:1979Natur}
{Mazets}, E.~P., {Golentskii}, S.~V., {Ilinskii}, V.~N., {Aptekar}, R.~L., \&
  {Guryan}, I.~A. 1979, \nat, 282, 587, \dodoi{10.1038/282587a0}

\bibitem[{{Meegan} {et~al.}(2009){Meegan}, {Lichti}, {Bhat}, {Bissaldi},
  {Briggs}, {Connaughton}, {Diehl}, {Fishman}, {Greiner}, {Hoover}, {van der
  Horst}, {von Kienlin}, {Kippen}, {Kouveliotou}, {McBreen}, {Paciesas},
  {Preece}, {Steinle}, {Wallace}, {Wilson}, \& {Wilson-Hodge}}]{Meegan:2009}
{Meegan}, C., {Lichti}, G., {Bhat}, P.~N., {et~al.} 2009, \apj, 702, 791,
  \dodoi{10.1088/0004-637X/702/1/791}

\bibitem[{{Nakar} {et~al.}(2005){Nakar}, {Piran}, \& {Sari}}]{Nakar:2005ApJ}
{Nakar}, E., {Piran}, T., \& {Sari}, R. 2005, \apj, 635, 516,
  \dodoi{10.1086/497296}

\bibitem[{{Negro} {et~al.}(2024){Negro}, {Younes}, {Wadiasingh}, {Burns},
  {Trigg}, \& {Baring}}]{Negro:2024FrASS1188953N}
{Negro}, M., {Younes}, G., {Wadiasingh}, Z., {et~al.} 2024, Frontiers in
  Astronomy and Space Sciences, 11, 1388953, \dodoi{10.3389/fspas.2024.1388953}

\bibitem[{{Paliya} {et~al.}(2019){Paliya}, {Dom{\'\i}nguez}, {Ajello},
  {Franckowiak}, \& {Hartmann}}]{Paliya:2019ApJ}
{Paliya}, V.~S., {Dom{\'\i}nguez}, A., {Ajello}, M., {Franckowiak}, A., \&
  {Hartmann}, D. 2019, \apjl, 882, L3, \dodoi{10.3847/2041-8213/ab398a}

\bibitem[{{Palmer} {et~al.}(2005){Palmer}, {Barthelmy}, {Gehrels}, {Kippen},
  {Cayton}, {Kouveliotou}, {Eichler}, {Wijers}, {Woods}, {Granot}, {Lyubarsky},
  {Ramirez-Ruiz}, {Barbier}, {Chester}, {Cummings}, {Fenimore}, {Finger},
  {Gaensler}, {Hullinger}, {Krimm}, {Markwardt}, {Nousek}, {Parsons}, {Patel},
  {Sakamoto}, {Sato}, {Suzuki}, \& {Tueller}}]{Palmer:2005Natur}
{Palmer}, D.~M., {Barthelmy}, S., {Gehrels}, N., {et~al.} 2005, \nat, 434,
  1107, \dodoi{10.1038/nature03525}

\bibitem[{{Preece} {et~al.}(1998){Preece}, {Briggs}, {Mallozzi}, {Pendleton},
  {Paciesas}, \& {Band}}]{Preece:1998}
{Preece}, R.~D., {Briggs}, M.~S., {Mallozzi}, R.~S., {et~al.} 1998, \apjl, 506,
  L23, \dodoi{10.1086/311644}

\bibitem[{{Principe} {et~al.}(2023){Principe}, {Di Venere}, {Negro}, {Di
  Lalla}, {Omodei}, {Di Tria}, {Mazziotta}, \& {Longo}}]{Principe:2023}
{Principe}, G., {Di Venere}, L., {Negro}, M., {et~al.} 2023, \aap, 675, A99,
  \dodoi{10.1051/0004-6361/202346492}

\bibitem[{{Svinkin} {et~al.}(2021){Svinkin}, {Frederiks}, {Hurley}, {Aptekar},
  {Golenetskii}, {Lysenko}, {Ridnaia}, {Tsvetkova}, {Ulanov}, {Cline},
  {Mitrofanov}, {Golovin}, {Kozyrev}, {Litvak}, {Sanin}, {Goldstein}, {Briggs},
  {Wilson-Hodge}, {von Kienlin}, {Zhang}, {Rau}, {Savchenko}, {Bozzo},
  {Ferrigno}, {Ubertini}, {Bazzano}, {Rodi}, {Barthelmy}, {Cummings}, {Krimm},
  {Palmer}, {Boynton}, {Fellows}, {Harshman}, {Enos}, \&
  {Starr}}]{Svinkin:2021Natur}
{Svinkin}, D., {Frederiks}, D., {Hurley}, K., {et~al.} 2021, \nat, 589, 211,
  \dodoi{10.1038/s41586-020-03076-9}

\bibitem[{{Trigg} {et~al.}(2024){Trigg}, {Burns}, {Roberts}, {Negro},
  {Svinkin}, {Baring}, {Wadiasingh}, {Christensen}, {Andreoni}, {Briggs}, {Di
  Lalla}, {Frederiks}, {Lipunov}, {Omodei}, {Ridnaia}, {Veres}, \&
  {Lysenko}}]{Trigg:2024}
{Trigg}, A.~C., {Burns}, E., {Roberts}, O.~J., {et~al.} 2024, \aap, 687, A173,
  \dodoi{10.1051/0004-6361/202348858}

\bibitem[{{Trigg} {et~al.}(2025){Trigg}, {Stewart}, {Van Kooten}, {Burns},
  {Baring}, {Frederiks}, {Huppenkothen}, {O'Connor}, {Roberts}, {Wadiasingh},
  {Younes}, {Bhat}, {Briggs}, {Busmann}, {Goldstein}, {Gruen}, {Hu},
  {Kouveliotou}, {Negro}, {Palmese}, {Riffeser}, {Scotton}, {Svinkin}, {Veres},
  \& {Z{\"o}ller}}]{Trigg:2025M82}
{Trigg}, A.~C., {Stewart}, R., {Van Kooten}, A., {et~al.} 2025, \aap, 694,
  A323, \dodoi{10.1051/0004-6361/202452268}

\bibitem[{{Trigg} {et~al.}(2026){Trigg}, {Burns}, {Negro}, {Bala}, {Bhat},
  {Cleveland}, {Frederiks}, {Goldstein}, {Hristov}, {Kocevski}, {Di Lalla},
  {Lesage}, {Mailyan}, {Neights}, {Omodei}, {Roberts}, {Scotton}, {Svinkin}, \&
  {Wood}}]{Trigg:2025}
{Trigg}, A.~C., {Burns}, E., {Negro}, M., {et~al.} 2026, \aap, 710, A139,
  \dodoi{10.1051/0004-6361/202558217}

\bibitem[{{Vianello}(2016)}]{Vianello2016gtburst}
{Vianello}, G. 2016, {gtburst: Release for Zenodo}, 02-01-03,  Zenodo,
  \dodoi{10.5281/zenodo.59783}

\bibitem[{{Xing} {et~al.}(2024){Xing}, {Yu}, {Yan}, {Zhang}, \&
  {Zhang}}]{Xing:2024}
{Xing}, Y., {Yu}, W., {Yan}, Z., {Zhang}, X., \& {Zhang}, B. 2024, arXiv
  e-prints, arXiv:2411.06996, \dodoi{10.48550/arXiv.2411.06996}

\bibitem[{{Yang} {et~al.}(2020){Yang}, {Chand}, {Zhang}, {Yang}, {Zou}, {Yang},
  {Zhao}, {Shao}, {Xiong}, {Luo}, {Li}, {Xiao}, {Li}, {Liu}, {Joshi}, {Sharma},
  {Chakraborty}, {Li}, \& {Zhang}}]{Yang:2020ApJ}
{Yang}, J., {Chand}, V., {Zhang}, B.-B., {et~al.} 2020, \apj, 899, 106,
  \dodoi{10.3847/1538-4357/aba745}

\bibitem[{{Zhang} {et~al.}(2020){Zhang}, {Liu}, {Zhong}, \&
  {Wang}}]{Zhang:2020ApJ}
{Zhang}, H.-M., {Liu}, R.-Y., {Zhong}, S.-Q., \& {Wang}, X.-Y. 2020, \apjl,
  903, L32, \dodoi{10.3847/2041-8213/abc2c9}

\end{thebibliography}

\end{document}